\documentclass[twocolumn,aps,pra,10pt]{revtex4-2}
\usepackage{amssymb}
\usepackage{amsfonts}
\usepackage{amsmath}
\usepackage{amsxtra}
\usepackage{amscd}
\usepackage{amsthm}
\usepackage{graphicx}
\usepackage{epsf}
\usepackage{bbold}
\usepackage{xcolor}
\usepackage{array}

\newcommand{\PreserveBackslash}[1]{\let\temp=\\#1\let\\=\temp}
\newcolumntype{C}[1]{>{\PreserveBackslash\centering}p{#1}}
\newcolumntype{R}[1]{>{\PreserveBackslash\raggedleft}p{#1}}
\newcolumntype{L}[1]{>{\PreserveBackslash\raggedright}p{#1}}

\begin{document}

\title{One-dimensional $\mathbb{Z}_4$ topological superconductor}

\author{Max Tymczyszyn}
\author{Edward McCann}
 \email{ed.mccann@lancaster.ac.uk}
\affiliation{Department of Physics, Lancaster University, Lancaster, LA1 4YB, United Kingdom}

\begin{abstract}
We describe the mean-field model of a one-dimensional topological superconductor with two orbitals per unit cell. Time-reversal symmetry is absent, but a nonsymmorphic symmetry, involving a translation by a fraction of the unit cell, mimics the role of time-reversal symmetry. As a result, the topological superconductor has $\mathbb{Z}_4$ topological phases, two which support Majorana bound states and two which do not, in agreement with a prediction based on K-theory classification [K. Shiozaki {\em et al.},
Phys.\ Rev.\ B {\bf 93}, 195413 (2016)].
As with the Kitaev chain, the presence of Majorana bound states gives rise to the $4\pi$-periodic Josephson effect.
A random matrix with nonsymmorphic time-reversal symmetry may be block diagonalized, and every individual block has time-reversal symmetry described by one of the Gaussian orthogonal, unitary or symplectic ensembles.
We show how this is manifested in the energy level statistics of a random system in the $\mathbb{Z}_4$ class as the spatial period of the nonsymmorphic symmetry is varied from much less than to of the order of the system size.
\end{abstract}

\maketitle

\begin{figure}[t]
\includegraphics[scale=0.18]{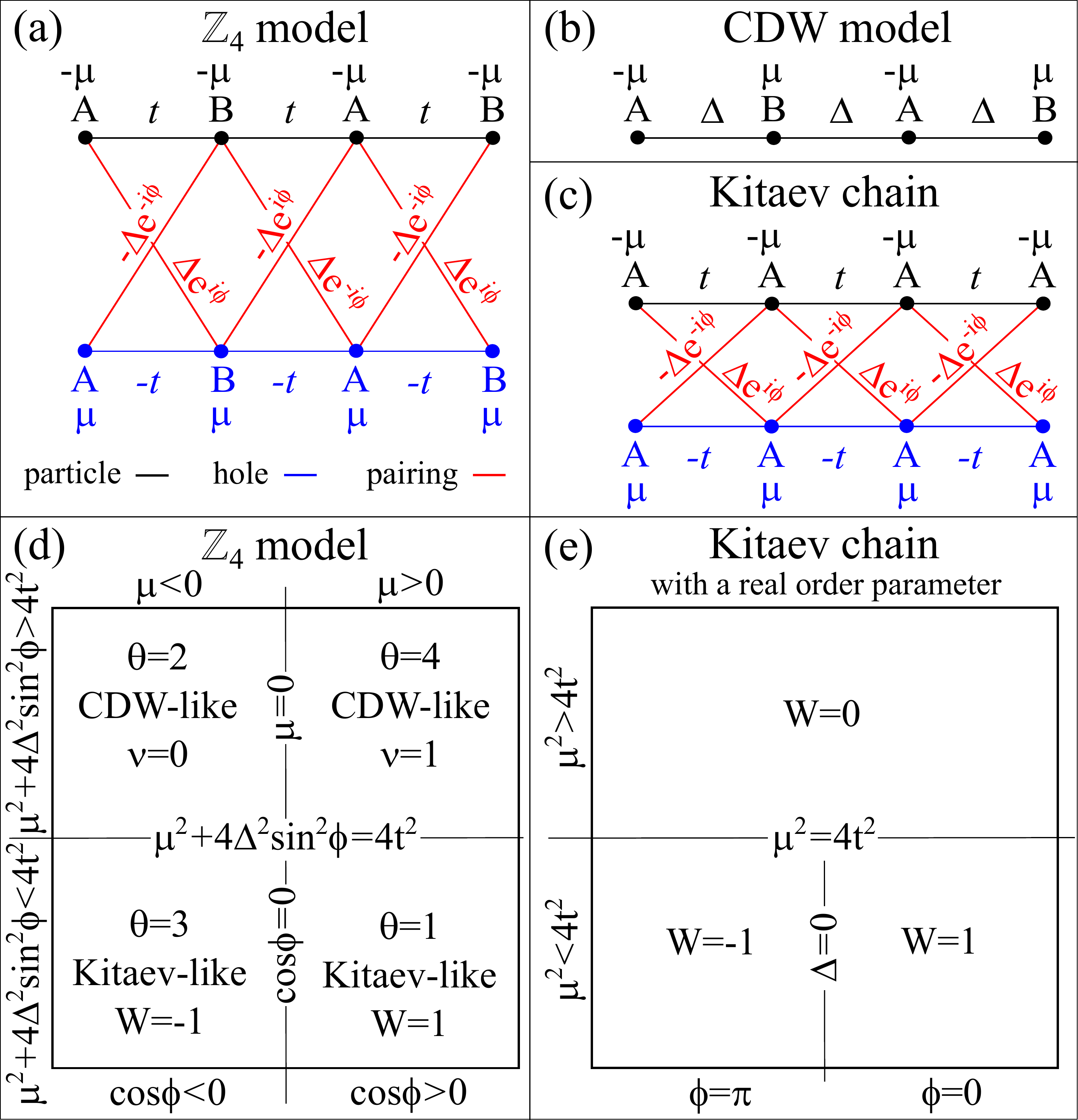}
\caption{(a) The $\mathbb{Z}_4$ model shown in the Bogoliubov de Gennes (BdG) representation with two orbitals $A$ and $B$ per cell, and particle-hole chains with onsite energies $\pm \mu$, nearest-neighbor hoppings $\pm t$, and superconducting couplings of magnitude $\Delta$ and phase $\phi$. (b) The noninteracting CDW model where $\Delta$ acts as nearest-neighbor hopping. (c) The Kitaev chain in the BdG representation.
In (a) and (c), the particle (hole) branch is shown in black (blue), and the superconducting pairing is shown in red.
Phase diagram of (d) the $\mathbb{Z}_4$ topological superconductor with phases $\theta = 1,2,3,4$ and (e) the Kitaev chain with a real order parameter. In (d), each phase is labeled using a phase of a two-band model.
The CDW model has two phases denoted $\nu = 0$ and $\nu =1$, and the Kitaev chain with nearest-neighbor hopping has three possible winding numbers $W = 0$, $1$, and $-1$.
For (d) and (e), in the designation of the Kitaev chain phases, we assume $t>0$.}
\label{z4fig1}
\end{figure}

The ten-fold way classification of topological insulators and superconductors~\cite{altland97,schnyder08,kitaev09,ryu10,chiu16} is based on the presence or absence of three nonspatial symmetries which act upon the degrees of freedom within a unit cell: Time-reversal symmetry, particle-hole (or charge-conjugation) symmetry, and chiral symmetry.
Subsequently, this classification was generalized to describe crystalline symmetries including nonsymmorphic symmetries which incorporate a translation by a fraction of a unit cell~\cite{liu14,shiozaki14,young15,wang16,shiozaki16,varjas17,kruthoff17,herrera22}.
Recently, there has been interest in systems where some nonspatial symmetries are broken, but a nonsymmorphic (NS) symmetry mimics the effect of the absent nonspatial symmetry~\cite{mong10,fang15,shiozaki15,zhao16,yanase17,arkinstall17,otrokov19,gong19,zhang19,niu20,marques19,brzezicki20,allen22,yang22}. 
Examples include three-dimensional antiferromagnetic topological insulators with NS time-reversal symmetry~\cite{mong10,otrokov19,gong19,zhang19,niu20} and the 
charge-density wave (CDW) model which is a one-dimensional two-band tight-binding model with constant nearest-neighbor hopping strengths and alternating onsite energies~\cite{kivelson83,shiozaki15,brzezicki20,cayssol21,fuchs21,allen22,mccann23}, and possessing NS chiral symmetry.

In this paper we describe the mean-field model of a one-dimensional topological superconductor with two orbitals per unit cell, yielding a four component Hamiltonian in the Bogoliubov de Gennes (BdG) representation. It possesses particle-hole symmetry, NS time-reversal symmetry and NS chiral symmetry.
We find that it has a $\mathbb{Z}_4$ topological index, in agreement with the prediction of K-theory classification~\cite{shiozaki16}. This is unique for one dimension, where topological classes in the ten-fold way are described by either $\mathbb{Z}$ or $\mathbb{Z}_2$ indices~\cite{schnyder08,kitaev09,ryu10,chiu16}, although nontrivial topological phases, including $\mathbb{Z}_4$ phases~\cite{shiozaki16}, have been discussed in higher spatial dimensions with surface states referred to as an hourglass or M\"obius strip~\cite{shiozaki15,wang16,shiozaki16,ezawa16,yanase17,chang17,daido19,day23}.

Figure~\ref{z4fig1}(a) shows the model in the BdG representation where the magnitude $\Delta$ and phase $\phi$ of the superconducting order parameter are shown as couplings between particle-hole chains with onsite energies $\pm \mu$ and nearest-neighbor hoppings $\pm t$, where $\mu$ is the chemical potential.
The two orbitals per unit cell are denoted $\sigma = A$ or $\sigma = B$ which are sublattice labels for spinless fermions.
Related two band models are the noninteracting charge-density wave (CDW) model~\cite{kivelson83,shiozaki15,brzezicki20,cayssol21,fuchs21,allen22,mccann23}, Fig.~\ref{z4fig1}(b), and the Kitaev chain~\cite{kitaev01}, shown in the BdG representation in Fig.~\ref{z4fig1}(c).
For the $\mathbb{Z}_4$ model, Figure~\ref{z4fig1}(a), the intercell superconducting order parameter is the complex conjugate of the intracell order parameter, unlike the Kitaev model, Fig.~\ref{z4fig1}(c).

The phase diagram of the $\mathbb{Z}_4$ model is shown in Fig.~\ref{z4fig1}(d) with four phases distinguished by topological index $\theta = 1,2,3,4$. We label the four phases by relating them to phases of the two band models. For $\mu^2 + 4 \Delta^2 \sin^2 \phi < 4t^2$, the phases ($\theta = 1$ and $\theta = 3$) may be reduced to phases of the Kitaev chain with a real order parameter~\cite{supplementary} by setting $\phi = 0$ or $\phi = \pi$, respectively, corresponding to winding number $W=1$ or $W=-1$.
For comparison, the phase diagram of the Kitaev chain with nearest-neighbor coupling~\cite{spanslatt15} is shown in Fig.~\ref{z4fig1}(e).
For $\mu^2 + 4 \Delta^2 \sin^2 \phi > 4t^2$, the phases ($\theta = 4$ and $\theta = 2$) may be reduced to phases of the CDW model by setting $t = 0$ with $\mu > 0$ or $\mu < 0$, respectively, corresponding to $\mathbb{Z}_2$ topological index $\nu = 1$ or $\nu =0$~\cite{shiozaki15}. This correspondence occurs because, for $t = 0$, the phase $\phi$ may be gauged away~\cite{supplementary} and the superconducting chain in the BdG represention may be interpreted as a noninteracting model (the CDW model).

The mean-field Hamiltonian of the $\mathbb{Z}_4$ topological superconductor is given by
\begin{eqnarray}
\!\!\! H &=& - \mu \sum_{n,\sigma} c_{n,\sigma}^{\dagger} c_{n,\sigma}
+ t \sum_n \left( c_{n,A}^{\dagger} c_{n,B} + c_{n+1,A}^{\dagger} c_{n,B} + \mathrm{H.c.} \right) \nonumber \\
&&
+ \Delta \sum_n \Big( e^{i \phi} c_{n,A}^{\dagger} c_{n,B}^{\dagger} + e^{-i \phi} c_{n,B}^{\dagger} c_{n+1,A}^{\dagger} + \mathrm{H.c.} \Big) , \label{h1}
\end{eqnarray}
where $c_{n,\sigma}^{\dagger}$, $c_{n,\sigma}$ are creation and annihilation operators for orbital $\sigma = \{ A , B \}$ in unit cell $n$.
The Hamiltonian~(\ref{h1}) may be written in the BdG representation~\cite{tanaka12,alicea12,leijnse12,beenakker15,guo16,sato17} as $H = \frac{1}{2} \sum_k \Psi_k^{\dagger} {\cal H} (k) \Psi_k + \frac{1}{2} \sum_k \mathrm{tr} (\hat{h}(k))$ where
$\Psi_k = \begin{pmatrix} c_{k A} &  c_{k B} & c_{-k A}^{\dagger} & c_{-k B}^{\dagger} \end{pmatrix}^T$
for wave vector $k$ and the $4 \times 4$ Bloch Hamiltonian ${\cal H} (k)$~\cite{supplementary,bena09} is
\begin{eqnarray}
{\cal H} (k) &=& \begin{pmatrix}
\hat{h} (k) & \hat{\Delta} (k) \\
\hat{\Delta}^{\dagger} (k) & -\hat{h}^{\ast} (-k)
\end{pmatrix} , \label{bdg1} \\
\hat{h} (k) &=& \begin{pmatrix}
- \mu & 2 t \cos (ka/2) \\
2 t \cos (ka/2)  & -\mu
\end{pmatrix} , \label{hk0} \\
\hat{\Delta} (k) &=& \begin{pmatrix}
0 & 2i\Delta \sin (ka/2 + \phi) \\
2i\Delta \sin (ka/2 - \phi)  & 0
\end{pmatrix} . \label{gf0}
\end{eqnarray}
The Bloch Hamiltonian incorporates particle-hole symmetry as
\begin{eqnarray}
\text{particle-hole:} \qquad 
C^{\dagger} {\cal H}^{\ast} (k) C = - {\cal H} (-k) , \label{phs}
\end{eqnarray}
where $C = \tau_x \sigma_0$, and we use $\tau_0$  and $\tau_i$ for the identity and Pauli matrices in the particle-hole space, and $\sigma_0$ and $\sigma_i$ for the identity and Pauli matrices in the sublattice space, where $i = x,y,z$.
The Hamiltonian breaks time-reversal symmetry~\cite{haim19}, but the NS time-reversal symmetry acts as
\begin{eqnarray}
\text{NS\,\,time:} \qquad 
T^{\dagger} {\cal H}^{\ast} (k) T = {\cal H} (-k) , \label{nstime}
\end{eqnarray}
where $T = \tau_0 \sigma_x$.
The combination of particle-hole symmetry and NS time-reversal symmetry yields a NS chiral symmetry,
$S^{\dagger} {\cal H} (k) S = - {\cal H} (k)$,
where $S = \tau_x \sigma_x$.

With the symmetry constraints, the gap function may be written generically in sublattice space as
\begin{eqnarray}
\hat{\Delta} (k) &=& \begin{pmatrix}
i d_y (k) - d_x (k) & i d_z (k) + i \Delta_s (k) \\
i d_z (k) - i \Delta_s (k) & i d_y (k) + d_x (k)
\end{pmatrix} . \label{gf1}
\end{eqnarray}
where the $\Delta_s$, $d_x$, $d_y$, $d_z$ components are real, $\Delta_s (k) = \Delta_s (-k)$ has even parity and $\mathbf{d}(k) = ( d_x (k) , d_y (k) , d_z (k) )$ has odd parity $\mathbf{d}(k) = - \mathbf{d}(-k)$.
This is consistent with the constraint of the Pauli exclusion principle $\hat{\Delta} (-k) = - \hat{\Delta}^T (k)$.
Hence, we could include diagonal sublattice terms $d_x$, $d_y$, but they are not needed to describe the $\mathbb{Z}_4$ topology, and we restrict the discussion to the off-diagonal gap function~(\ref{gf0}).
The energy spectrum $E(k)$ of the BdG Hamiltonian~(\ref{bdg1}) has four bands given by
\begin{eqnarray*}
E^2 (k) &=& \mu^2 + f^2 + d_z^2 + \Delta_s^2
\pm 2 \sqrt{\mu^2 f^2 + \Delta_s^2 ( f^2 + d_z^2)} , \\
f (k) &=& 2 t \cos (ka/2) , \\
\Delta_s (k) &=& 2 \Delta \sin \phi \cos (ka/2) , \\
d_z (k) &=& 2 \Delta \cos \phi \sin (ka/2) .
\end{eqnarray*}
Despite the mixed parity of the gap function~(\ref{gf0}), the presence of particle-hole symmetry and NS time-reversal symmetry ensure particle-hole symmetry of the energy spectrum and its evenness in $k$.
Furthermore, NS time-reversal symmetry imposes Kramer's degeneracy~\cite{zhao16,mccann23} with the two positive-energy bands being degenerate at the Brillouin zone edge, likewise the two negative-energy bands.
The degeneracy occurs at only one time-reversal invariant $k$ value (either $k=0$ or $k=\pi/a$) as opposed to both, which happens for symmorphic time-reversal symmetry with $T^2 = -1$~\cite{zhao16,mccann23}.
Despite the Kramer's degeneracy, there is generally a band gap between the positive-energy and negative-energy bands; parameter values leading to a vanishing band gap are indicated in the phase diagram, Fig.~\ref{z4fig1}(d).

\begin{figure}[t]
\includegraphics[scale=0.53]{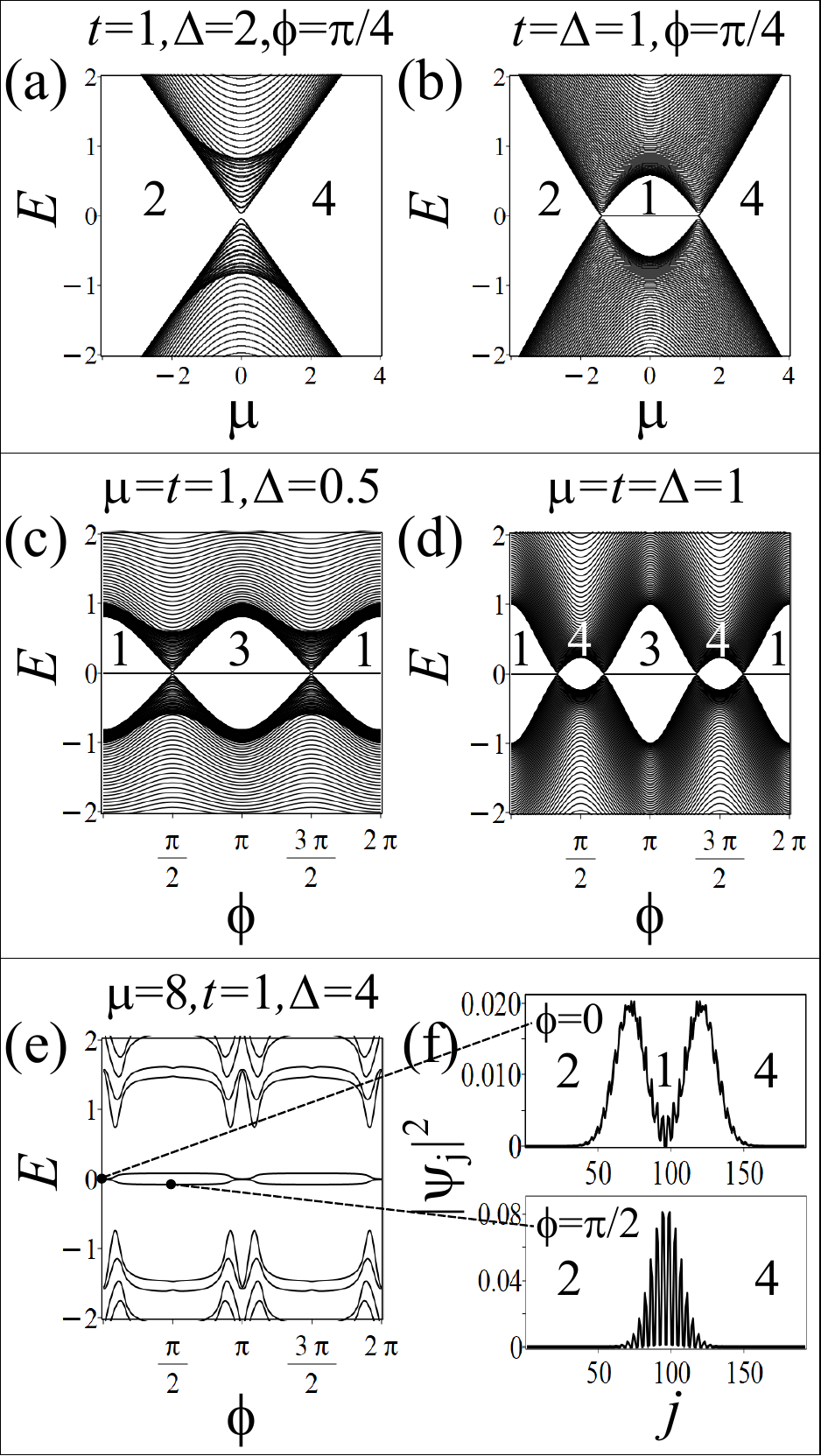}
\caption{Energy levels of the $\mathbb{Z}_4$ model in position space.
(a) and (b) are as a function of the chemical potential $\mu$ with (a) $\Delta = 2.0$, $\phi = \pi /4$, and (b) $\Delta = 1.0$, $\phi = \pi /4$.
(c) and (d) are as a function of the superconducting phase $\phi$ with (c) $\mu = 1.0$, $\Delta = 0.5$, and (d) $\mu = \Delta = 1.0$. 
(e) Energy levels as a function of $\phi$  for a system with a soliton in the onsite energies~(\ref{soliton}) of width $\zeta = 24$ at its center, for $\mu = 8.0$, $t=1.0$, $\Delta = 4.0$.
(f) The probability density $|\psi_j|^2$ per site $j = 1,2,\ldots , 192$ for the state at negative energy localized on the soliton for $\phi = 0$ (top) and $\phi = \pi/2$ (bottom).
Numbers in each plot show the phases $\theta$ of the $\mathbb{Z}_4$ model.
All plots are obtained by diagonalizing the BdG Hamiltonian in position space with open boundary conditions, $48$ unit cells and $t = 1.0$~\cite{bdgnote}.
}
\label{levelsfig1}
\end{figure}

The $\mathbb{Z}_4$ topological index $\theta$ has been defined~\cite{shiozaki16} as
\begin{eqnarray}
\theta &=& - \frac{2}{\pi} \arg \big\{ \mathrm{Pf} [ \sigma_x Z(\pi/a) ] \big\} \nonumber \\
&& \qquad + \frac{1}{\pi} \int_0^{\pi/a} dk \frac{\partial}{\partial k} \arg \{ \det [ \sigma_x Z(k) ] \} , \label{thetadef}
\end{eqnarray}
where $\theta$ is defined modulo $4$ and $\arg(z) = \arg(r e^{i \varphi}) = \varphi$ for $-\pi < \varphi \leq \pi$~\cite{thetanote}. For a $2 \times 2$ antisymmetric matrix, the Pfaffian is
$\mathrm{Pf} (i \alpha \sigma_y) = \alpha$.
The $2 \times 2$ matrix $Z(k) = X(k) + i Y(k)$ consists of a $2\pi$ periodic Hermitian part $X(k)$ and an aperiodic Hermitian part $Y(k)$ with $X (\pi /a) = X (-\pi /a)$ and $Y (\pi /a) = -Y (-\pi /a)$ such that the BdG Hamiltonian ${\cal H} (k)$ may be rotated by a unitary transformation to $\tilde{{\cal H}} (k) = \tilde{U}^{\dagger}{\cal H} (k) \tilde{U}$ where
\begin{eqnarray}
\tilde{{\cal H}} (k) = \begin{pmatrix}
X(k) & i Y(k) \\
-iY(k) & - X(k)
\end{pmatrix} .
\end{eqnarray}
For the BdG Hamiltonian~(\ref{bdg1}), we find that
\begin{eqnarray*}\!\!\!
Z(k) = \begin{pmatrix}
-\mu + 2i\Delta \sin (ka/2 + \phi)  & 2t\cos (ka/2) \\
-2t\cos (ka/2) & \mu - 2i\Delta \sin (ka/2 - \phi)
\end{pmatrix} ,
\end{eqnarray*}
and, substituting into Eq.~(\ref{thetadef}), this gives the values of $\theta$ shown in the phase diagram, Fig.~\ref{z4fig1}(d).
Note the presence of a discontinuity in the integral of Eq.~(\ref{thetadef}) in the case $\mu^2 + 4 \Delta^2 \sin^2 \phi < 4t^2$ and $\mu \cos \phi > 0$.

The presence of Majorana bound states is indicated by the $\mathbb{Z}_2$ Majorana number $M$~\cite{kitaev01,budich13,beenakker15,chiu16,guo16,sato17}, and we find $M = \mathrm{sgn} (\mu^2 + 4 \Delta^2 \sin^2 \phi - 4 t^2 )$~\cite{supplementary} indicating that the Kitaev-like phases, $\theta = 1, 3$, support Majorana bound states ($M=-1$) whereas the CDW-like phases, $\theta = 2, 4$, do not ($M=1$).
To explore the phase diagram, Fig.~\ref{z4fig1}(d), we plot the energy levels in position space  in Fig.~\ref{levelsfig1}.
Fig.~\ref{levelsfig1}(a) shows the energy levels as a function of $\mu$ for $\Delta^2 \sin^2 \phi > t^2$. There is a phase transition at $\mu = 0$ between the CDW-like phases $\theta = 2$ and $\theta = 4$ that has no analog in the Kitaev model.
For $\Delta^2 \sin^2 \phi < t^2$, Fig.~\ref{levelsfig1}(b), there are two phase transitions, from $\theta = 2$ to $\theta = 1$ at $\mu < 0$  and from $\theta = 1$ to $\theta = 4$ at $\mu > 0$. The Kitaev-like phase, $\theta = 1$, supports edge states at zero energy.
Fig.~\ref{levelsfig1}(c) shows the energy levels as a function of the superconducting phase $\phi$ with $\mu^2 + 4 \Delta^2 \sin^2 \phi < 4t^2$ for all $\phi$. Thus, there are transitions between the Kitaev-like phases $\theta = 1, 3$. In Fig.~\ref{levelsfig1}(d), there are also transitions across the boundary $\mu^2 + 4 \Delta^2 \sin^2 \phi = 4t^2$ to the CDW-like phase $\theta = 4$.
Note that the same plots for the Kitaev model would be independent of $\phi$ (being equal to the plots shown at $\phi = 0$ for all $\phi$) because $\phi$ can be gauged away in the Kitaev model.

The phase diagram can be explored further by considering the influence of a soliton which is a spatial texture separating different phases in position space.
In particular, we implement a texture in the onsite energies with the energy of an $A$ site in unit cell $n$ given by
\begin{eqnarray}
\epsilon_{n,A} = \mu \tanh \left( \frac{n - N/2 - 1/2}{\zeta} \right) , \label{soliton}
\end{eqnarray}
where $N$ is the number of unit cells and $\zeta$ is the soliton width in dimensionless units, i.e., measured in units of the lattice constant.
We assume that the onsite energies have the same magnitude within a unit cell~\cite{brzezicki20,allen22}.
Energy levels for a soliton located at the center of the system are plotted as a function of $\phi$ in Fig.~\ref{levelsfig1}(e).
For all $\phi$ values, there are two low-energy levels near zero energy; they appear in pairs owing to the particle-hole symmetry. The state of the lowest energy level of each pair is plotted in position space in Fig.~\ref{levelsfig1}(f).
The top panel, for $\phi = 0$, shows that this state is localized on two spatial regions where there is a boundary between $\theta = 2$ and $\theta = 1$, and a boundary between $\theta = 1$ and $\theta = 4$.
The bottom panel, for $\phi = \pi /2$, shows that this state is localized on only one spatial region where there is a boundary between $\theta = 2$ and $\theta = 4$. In the latter case, the energy of the localized state is not exactly at zero energy because the NS symmetry is only approximate in the presence of a soliton of finite width, as has been previously studied for the CDW model~\cite{brzezicki20,allen22}.

\begin{figure}[t]
\includegraphics[scale=0.5]{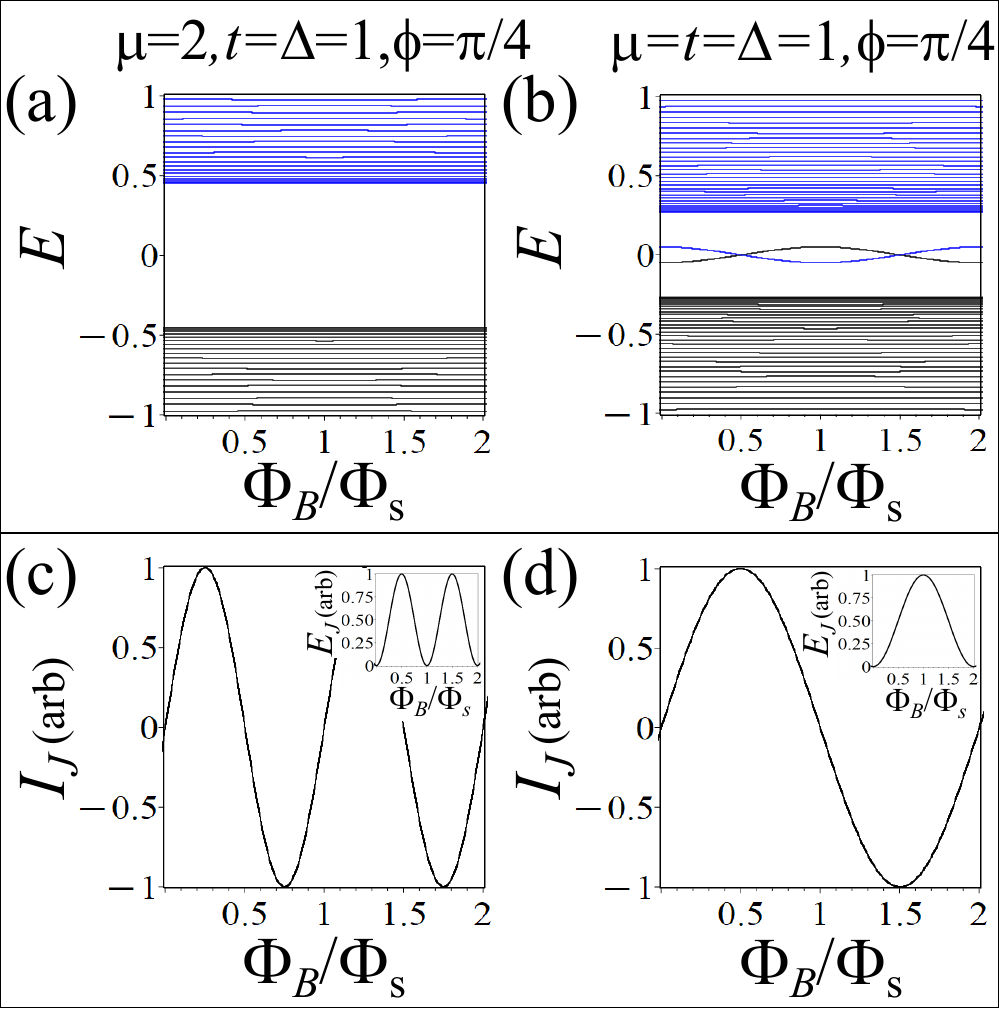}
\caption{Properties of a Josephson junction in a ring with phase $\theta = 4$ (left column with $\mu = 2.0$) and $\theta = 1$ (right column with $\mu =1.0$). All plots are as a function of the magnetic flux $\Phi_B$ where $\Phi_{\mathrm{s}} = h/2e$ is the superconducting flux quantum.
(a) and (b) show energy levels $E$ in position space.
(c), (d) show the current $I_{\mathrm{J}}$ (arbitrary units), with the corresponding system energy $E_{\mathrm{J}}$ in the insets.
The energy $E_{\mathrm{J}}$ is determined by including the negative energy levels at $\Phi_B = 0$ (black), excluding the positive ones (blue).
All plots are obtained by diagonalizing the BdG Hamiltonian in position space  with $48$ unit cells in the ring, $t = \Delta = 1.0$, $\phi = \pi/4$, and $t^{\prime} = 0.1$~\cite{bdgnote}.
}
\label{josephsonfig}
\end{figure}

As in the Kitaev chain~\cite{kitaev01,kwon04,alicea12,pientka13,beenakker15}, Majorana bound states in the $\mathbb{Z}_4$ model give rise to a $4\pi$-periodic Josephson junction. To illustrate this, we generalize a model of a weak link in a closed ring~\cite{pientka13} applied previously to the Kitaev chain.
For a ring with $N$ unit cells, the Hamiltonian~(\ref{h1}) is supplemented by a term describing hopping of magnitude $t^{\prime}$ across the weak link,
\begin{eqnarray}
\delta H = t^{\prime} c_{1,A}^{\dagger} c_{N,B} e^{-i \pi \Phi_B / \Phi_{\mathrm{s}}} + \mathrm{H.c.} ,
\end{eqnarray}
where $\Phi_B$ is the magnetic flux through the ring and $\Phi_{\mathrm{s}} = h/2e$ is the superconducting flux quantum.
We then diagonalize the BdG Hamiltonian in position space~\cite{supplementary} to obtain the energy levels $E$ and the energy of the junction $E_{\mathrm{J}}$ determined by summing the energy levels below zero energy. We calculate the derivative $dE_{\mathrm{J}}/d\Phi_B$ which is proportional to the Josephson current $I_{\mathrm{J}} = (2\pi / \Phi_{\mathrm{s}}) dE_{\mathrm{J}}/d\Phi_B$.
Fig.~\ref{josephsonfig}(a) shows the energy levels for the CDW-like phase $\theta = 4$ where Majorana bound states are absent. Although the levels appear to have little dependence on $\Phi_B$, Fig.~\ref{josephsonfig}(c) shows that the energy  $E_{\mathrm{J}}$ and current $I_{\mathrm{J}}$ are periodic in $\Phi_B$ with period $\Phi_{\mathrm{s}}$ as expected for a phase with tunneling of Cooper pairs and no Majorana bound states.

Fig.~\ref{josephsonfig}(b) shows the energy levels for the Kitaev-like phase $\theta = 1$ which shows two subgap states. We assume that the level crossings of these states (at $\Phi_B = \Phi_{\mathrm{s}}/2$ and $\Phi_B = 3\Phi_{\mathrm{s}}/2$) are protected by conservation of fermion parity~\cite{kitaev01,kwon04,alicea12,pientka13,beenakker15} so that these two levels have a period of $2\Phi_{\mathrm{s}}$, i.e., the single particle flux quantum. Hence, we do not determine the thermodynamic ground state energy, but include the level which is below zero energy at $\Phi_B = 0$ for all $\Phi_B$ in the calculation of $E_{\mathrm{J}}$ (shown as a black line in Fig.~\ref{josephsonfig}(b)), and omit the energy of the level above zero energy at $\Phi_B = 0$ for all $\Phi_B$ (a blue line in Fig.~\ref{josephsonfig}(b)). Fig.~\ref{josephsonfig}(d) shows that the resulting energy  $E_{\mathrm{J}}$ and current $I_{\mathrm{J}}$ are periodic in $\Phi_B$ with period $2\Phi_{\mathrm{s}}$ as expected for single-electron tunneling mediated by Majorana bound states~\cite{kitaev01,kwon04,alicea12,pientka13,beenakker15}.

\begin{figure}[t]
\includegraphics[scale=0.5]{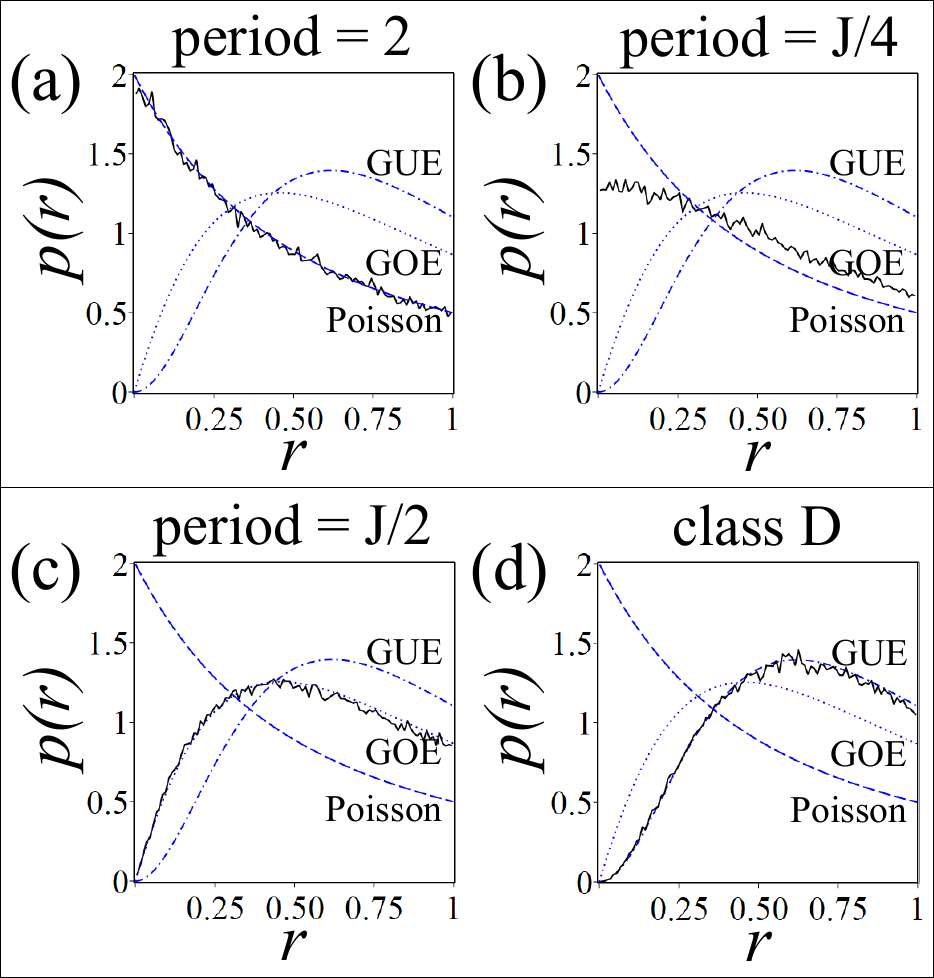}
\caption{
Distribution of the ratio of consecutive level spacings $p (r)$ for random matrices in the same class as the $\mathbb{Z}_4$ model with NS time-reversal symmetry of different spatial periods and particle-hole symmetry. (a) period equal to $2$ orbitals, (b) period equal to $J/4$ where $J$ is the total number of orbitals, (c) period equal to $J/2$, (d) class D where NS time-reversal symmetry is absent.
In all plots, black solid lines shown numerical data, blue dashed lines show the prediction of Poisson statistics, blue dotted lines show the prediction of the GOE ensemble~(\ref{pr}), and blue dot-dashed lines show the prediction of the GUE ensemble~(\ref{pr}).
We numerically diagonalized $J \times J$ random matrices with $J = 4000$, where every independent real variable is taken from a standard normal distribution, and averaged over an ensemble of $100$ matrices. Only positive energy levels were used for all plots, and twofold degeneracy was neglected (for a degenerate pair of levels, only one was included).
}
\label{rmtfig}
\end{figure}

The $\mathbb{Z}_4$ symmetry class is not robust in the presence of spatial disorder because of the constraints of translational invariance inherent to the NS symmetries~\cite{shiozaki16,allen22,supplementary}. However, when the NS symmetries are broken by disorder, particle-hole symmetry remains robust and the system falls into the D symmetry class~\cite{altland97,kitaev01,schnyder08,kitaev09,ryu10,budich13,beenakker15,chiu16,guo16,sato17} with $\mathbb{Z}_2$ Majorana number $M = \mathrm{sgn} (\mu^2 + 4 \Delta^2 \sin^2 \phi - 4 t^2 )$~\cite{supplementary}. This is distinct from the Kitaev chain, for which $M = \mathrm{sgn} (\mu^2 - 4 t^2 )$~\cite{kitaev01,budich13,beenakker15,chiu16,guo16,sato17}.

Symmetry classes generally exhibit different topological indices and numbers of zero-energy states~\cite{schnyder08,kitaev09,ryu10,chiu16}, but also different level statistics in the bulk of the energy spectrum as determined by time-reversal symmetry~\cite{mehta90,altland97,guhr98,beenakker15}.
A random matrix consistent with NS symmetry has a highly-constrained form and does not describe a system with arbitrary spatial disorder but a system with sample-to-sample parametric variations~\cite{allen22}: Parameters must be spatially uniform within a given sample, but they make take different values between members of an ensemble. Thus, a $J \times J$ random matrix consists of only $\sim J$ independent real parameters~\cite{allen22} instead of $\sim J^2$ in the absence of NS symmetry.

Specifically, we consider the level statistics of a large $J \times J$ Hermitian matrix ${\mathcal H}$ with random elements~\cite{supplementary} satisfying the constraints of particle-hole symmetry ${\mathcal C}^{\dagger} {\mathcal H}^{\ast} {\mathcal C} = -{\mathcal H}$ and NS time-reversal symmetry ${\mathcal T}^{\dagger} {\mathcal H}^{\ast} {\mathcal T} = {\mathcal H}$. In particular, ${\mathcal C} = \mathrm{diag} (\sigma_x , \sigma_x , \ldots , \sigma_x)$ is block diagonal and ${\mathcal T}$ involves a translation by two orbitals,
\begin{eqnarray}
{\mathcal T} =
\begin{pmatrix}
0 & 0 & 1 & 0 & 0 & 0 & \hdots \\
0 & 0 & 0 & 1 & 0 & 0 & \hdots \\
0 & 0 & 0 & 0 & 1 & 0 & \hdots \\
0 & 0 & 0 & 0 & 0 & 1 & \hdots \\
\vdots & \vdots & \vdots & \vdots & \vdots & \vdots & \ddots \\
1 & 0 & 0 & 0 & 0 & 0 & \hdots \\
0 & 1 & 0 & 0 & 0 & 0 & \hdots
\end{pmatrix} . \label{tposition}
\end{eqnarray}
With this representation, ${\mathcal T}^2 \neq \pm 1$, unlike conventional time-reversal symmetry~\cite{ryu10,chiu16}, meaning it can be applied twice to give a unitary transformation which is simply translation by a lattice constant (of four orbitals). This unitary transformation can then block diagonalize the random matrix into $4 \times 4$ blocks representing individual $k$ values.

Each block has a generic form~(\ref{bdg1}) satisfying particle-hole symmetry with gap function $\hat{\Delta} (k)$ as given in Eq.~(\ref{gf1}) and noninteracting part $\hat{h}(k) = {\boldsymbol\sigma} \cdot {\bf h} (k)$ with ${\boldsymbol\sigma} = (\sigma_0,\sigma_x , \sigma_y , \sigma_z )$ and ${\bf h} = (h_0,h_x,h_y,h_z)$, where $h_0(k)$, $h_x(k)$, $h_y(k)$,  are even functions of $k$ and $h_z(k)$ is odd.
Each $4 \times 4$ block has a fixed value of $k$ and, in addition to particle-hole symmetry, it satisfies a different type of time-reversal symmetry depending on the $k$ value. Owing to the unusual Kramer's degeneracy associated with NS time-reversal symmetry~\cite{zhao16,mccann23}, the block corresponding to $k = 0$ has time-reversal symmetry with ${\mathcal T}^2 = 1$ (denoted by $\beta = 1$) whereas the block corresponding to $k = \pi / a$ has ${\mathcal T}^2 = -1$ ($\beta = 4$)~\cite{supplementary}.
The remaining blocks have no time-reversal symmetry individually ($\beta = 2$) but they appear in time-reversal symmetric pairs corresponding to $\pm k$, and the levels arising from one block in a pair are degenerate with those from the other block in the pair.

We confirm this picture by considering $J \times J$ Hermitian random matrices satisfying the symmetry constraints with each independent real parameter in the matrix taken from the standard normal distribution (with mean zero and variance of $1$). We numerically determine the eigenvalues $E_n$, the level spacing $s_n = (E_{n+1} - E_n)$, the ratio of consecutive level spacings $r_n = \text{min}\{ s_n , s_{n-1} \}/\text{max}\{ s_n , s_{n-1} \}$~\cite{oganesyan07,atas13},
such that $0 \leq r_n \leq 1$, and we determine the distribution of ratios $p (r)$.
We compare our numerical results with predictions for the  Poisson distribution, the Gaussian orthogonal ensemble (GOE) with $\beta = 1$, and the Gaussian unitary ensemble (GUE) with $\beta = 2$~\cite{mehta90,guhr98}.
For the ratio distribution~\cite{oganesyan07,atas13},
\begin{eqnarray}
p(r) =
\begin{cases}
\frac{2}{(1+r)^2} , & \text{Poisson} \\
\frac{27}{4} \frac{(r+r^2)}{(1+r+r^2)^{5/2}} , & \text{GOE}\,\,(\beta = 1)\\
\frac{81\sqrt{3}}{2\pi} \frac{(r+r^2)^2}{(1+r+r^2)^{4}} . & \text{GUE}\,\,(\beta = 2)
\end{cases} \label{pr}
\end{eqnarray}

Figure~\ref{rmtfig}(a) shows the numerically determined ratio distribution $p(r)$ for a $J \times J$ matrix with $J \gg 2$, where $2$ is the period of the NS time-reversal symmetry. The distribution appears to follow the Poisson distribution.
Although most $4 \times 4$ blocks have $\beta = 2$ level statistics, the effect of including many blocks gives a spectrum with almost independent energy levels.
While keeping the matrix order $J$ fixed, we increase the period of the NS time-reversal symmetry~(\ref{tposition}), increasing the size of each block and reducing the number of blocks.
Figure~\ref{rmtfig}(b) shows the numerically determined ratio distribution $p(r)$ for a system where the period is equal to $J/4$. In this case, there are only two $J/2 \times J/2$ blocks, one corresponds to $k=0$ with $\beta = 1$ statistics and one corresponds to $k=\pi/a$ with $\beta = 4$ statistics. The combination of these gives a spectrum that appears to be a crossover between Poisson and Gaussian.
Figure~\ref{rmtfig}(c) shows $p(r)$ for a system where the period is equal to $J/2$.
Now the distribution follows the GOE distribution because there is only one $J \times J$ block, corresponding to $k=0$ with $\beta = 1$ statistics.
Finally, Fig.~\ref{rmtfig}(d) shows a system with class D level statistics which follows the GUE distribution ($\beta = 2$); this is equivalent to a system in which the NS time-reversal symmetry is absent (or, equivalently, it can be viewed as having period $J$).

Different methods of creating the Kitaev chain, or similar systems, have been investigated, usually with a conventional $s$-wave superconductor in proximity to a system with strong spin-orbit coupling and/or magnetic field. 
Examples include hybrid superconductor-semiconductor nanowires~\cite{lutchyn10,oreg10,leijnse12,prada20},
a superconductor near to an array of magnetic atoms or nanoparticles~\cite{choy11,nadj-perge13} or in a spiraling magnetic field~\cite{martin12,kjaergaard12}, planar semiconductor-superconductor heterostructures~\cite{hell17,pientka17},
a chain of quantum dots with spin-orbit coupling and connected to superconductors~\cite{sau12,leijnse12b,fulga13,tsintzis24},
and spin-orbit-coupled wires, each proximity coupled to superconductors with different phases~\cite{lesser21a,lesser21b}.
The latter two platforms provide the option of tuning the phase $\phi$ of the order parameter on each site separately and, with them, it may be possible to engineer a system described by the $\mathbb{Z}_4$ Hamiltonian~(\ref{h1}).

To summarize, we have identified the mean-field model of a topological superconductor with two orbitals per cell in one dimension belonging to a $\mathbb{Z}_4$ class.
In the phase diagram, Fig.~\ref{z4fig1}(d), two phases have no Majorana bound states and they may be related to the phases of a $\mathbb{Z}_2$ noninteracting model, the CDW model. Two phases, which support Majorana bound states, may be related to topological phases of the Kitaev chain and, like the Kitaev chain, the presence of Majorana bound states gives rise to the $4\pi$-periodic Josephson effect.
A random matrix with NS time-reversal symmetry may be block diagonalized into many blocks, one with GOE level statistics, one with GSE (Gaussian symplectic ensemble) statistics, and the others appearing as time-reversal symmetric pairs. For such a random matrix describing a system size $J$ much greater than the period of the NS time-reversal symmetry, the level statistics will appear to be uncorrelated as described by the Poisson distribution. As the period approaches the system size $J$, the statistics will crossover to Gaussian, with GOE statistics when the period is equal to $J/2$ and, finally, GUE statistics when the NS time-reversal symmetry is absent.

All relevant data presented in this paper can be accessed~\cite{datanote}.

\begin{acknowledgments}
The authors thank C. Y. Leung, A. Romito, K. Shiozaki, D. Varjas, and J. H. Winter for helpful discussions.
\end{acknowledgments}

\onecolumngrid
\clearpage
\begin{center}
\textbf{\large Supplementary material: One-dimensional $\mathbb{Z}_4$ topological superconductor}
\end{center}

\setcounter{equation}{0}
\setcounter{figure}{0}
\setcounter{table}{0}
\setcounter{page}{1}
\makeatletter
\renewcommand{\theequation}{S\arabic{equation}}
\renewcommand{\thefigure}{S\arabic{figure}}
\renewcommand{\bibnumfmt}[1]{[S#1]}
\renewcommand{\citenumfont}[1]{S#1}

\section{Bogoliubov de Gennes representation in position space}

\subsection{Hamiltonian}

The mean-field Hamiltonian of the $\mathbb{Z}_4$ topological superconductor is given by
\begin{eqnarray}
H = - \mu \sum_{n,\sigma} c_{n,\sigma}^{\dagger} c_{n,\sigma}
+ t \sum_n \left( c_{n,\alpha}^{\dagger} c_{n,\beta} + c_{n+1,\alpha}^{\dagger} c_{n,\beta} + \mathrm{H.c.} \right)
+ \Delta \sum_n \Big( e^{-i \phi} c_{n,\beta} c_{n,\alpha} + e^{i \phi} c_{n+1,\alpha} c_{n,\beta} + \mathrm{H.c.} \Big) , \label{sh1}
\end{eqnarray}
where $c_{n,\sigma}^{\dagger}$, $c_{n,\sigma}$ are creation and annihilation operators for orbital $\sigma$ in unit cell $n$. 
Parameter $\mu$ is the chemical potential, $t$ is nearest-neighbor hopping, and $\Delta$ and $\phi$ are the magnitude and phase of the superconducting order parameter.
The two orbitals per unit cell are denoted $\sigma = \alpha$ or $\sigma = \beta$ which could be spin labels $\uparrow, \downarrow$ or, for spinless fermions, sublattice labels $A, B$.

To give an example of the Bogoliubov de Gennes (BdG) representation of Eq.~(\ref{sh1}) in position space, we consider $N=3$ unit cells with open boundary conditions. The expressions may be easily generalized to other $N$ values. $N=3$ has six orbitals, or twelve orbitals in the BdG representation giving a $12 \times 12$ matrix in position space.
In particular, we write $H = (1/2) \Psi^{\dagger} {\cal H} \Psi - N \mu$,
where
\begin{eqnarray}
\Psi^{\dagger} = \begin{pmatrix}
c_{1,\alpha}^{\dagger} & c_{1,\alpha} & c_{1,\beta}^{\dagger} & c_{1,\beta} & c_{2,\alpha}^{\dagger} & c_{2,\alpha} & c_{2,\beta}^{\dagger} & c_{2,\beta} & c_{3,\alpha}^{\dagger} & c_{3,\alpha} & c_{3,\beta}^{\dagger} & c_{3,\beta}
\end{pmatrix} ; \qquad \qquad
\Psi = \begin{pmatrix}
c_{1,\alpha} \\
c_{1,\alpha}^{\dagger} \\
c_{1,\beta} \\
c_{1,\beta}^{\dagger} \\
c_{2,\alpha} \\
c_{2,\alpha}^{\dagger} \\
c_{2,\beta} \\
c_{2,\beta}^{\dagger} \\
c_{3,\alpha} \\
c_{3,\alpha}^{\dagger} \\
c_{3,\beta} \\
c_{3,\beta}^{\dagger}
\end{pmatrix} , \label{kitaevbasis}
\end{eqnarray}
\begin{eqnarray}
\!\!\!{\cal H} =
\begin{pmatrix}
-\mu & 0 & t & \Delta e^{i\phi} & 0 & 0 & 0 & 0 & 0 & 0 & 0 & 0 \\
0 & \mu & -\Delta e^{-i\phi} & -t & 0 & 0 & 0 & 0 & 0 & 0 & 0 & 0 \\
t & -\Delta e^{i\phi} & -\mu & 0 & t & \Delta e^{-i\phi} & 0 & 0 & 0 & 0 & 0 & 0 \\
\Delta e^{-i\phi} & -t & 0 & \mu & -\Delta e^{i\phi} & -t & 0 & 0 & 0 & 0 & 0 & 0 \\
0 & 0 & t & -\Delta e^{-i\phi} & -\mu & 0 & t & \Delta e^{i\phi} & 0 & 0 & 0 & 0 \\
0 & 0 & \Delta e^{i\phi} & -t & 0 & \mu & -\Delta e^{-i\phi} & -t & 0 & 0 & 0 & 0 \\
0 & 0 & 0 & 0 & t & -\Delta e^{i\phi} & -\mu & 0 & t & \Delta e^{-i\phi} & 0 & 0 \\
0 & 0 & 0 & 0 & \Delta e^{-i\phi} & -t & 0 & \mu & -\Delta e^{i\phi} & -t & 0 & 0 \\
0 & 0 & 0 & 0 & 0 & 0 & t & -\Delta e^{-i\phi} & -\mu & 0 & t & \Delta e^{i\phi} \\
0 & 0 & 0 & 0 & 0 & 0 & \Delta e^{i\phi} & -t & 0 & \mu & -\Delta e^{-i\phi} & -t \\
0 & 0 & 0 & 0 & 0 & 0 & 0 & 0 & t & -\Delta e^{i\phi} & -\mu & 0 \\
0 & 0 & 0 & 0 & 0 & 0 & 0 & 0 & \Delta e^{-i\phi} & -t & 0 & \mu
\end{pmatrix} . \label{hpos}
\end{eqnarray}

\subsection{Symmetry operators}

Particle-hole symmetry (charge conjugation) is ${\mathcal C}^{\dagger} {\mathcal H}^{\ast} {\mathcal C} = -{\mathcal H}$. For the position space Hamiltonian~(\ref{hpos}),
\begin{eqnarray}
{\mathcal C} =
\begin{pmatrix}
0 & 1 & 0 & 0 & 0 & 0 & 0 & 0 & 0 & 0 & 0 & 0 \\
1 & 0 & 0 & 0 & 0 & 0 & 0 & 0 & 0 & 0 & 0 & 0 \\
0 & 0 & 0 & 1 & 0 & 0 & 0 & 0 & 0 & 0 & 0 & 0 \\
0 & 0 & 1 & 0 & 0 & 0 & 0 & 0 & 0 & 0 & 0 & 0 \\
0 & 0 & 0 & 0 & 0 & 1 & 0 & 0 & 0 & 0 & 0 & 0 \\
0 & 0 & 0 & 0 & 1 & 0 & 0 & 0 & 0 & 0 & 0 & 0 \\
0 & 0 & 0 & 0 & 0 & 0 & 0 & 1 & 0 & 0 & 0 & 0 \\
0 & 0 & 0 & 0 & 0 & 0 & 1 & 0 & 0 & 0 & 0 & 0 \\
0 & 0 & 0 & 0 & 0 & 0 & 0 & 0 & 0 & 1 & 0 & 0 \\
0 & 0 & 0 & 0 & 0 & 0 & 0 & 0 & 1 & 0 & 0 & 0 \\
0 & 0 & 0 & 0 & 0 & 0 & 0 & 0 & 0 & 0 & 0 & 1 \\
0 & 0 & 0 & 0 & 0 & 0 & 0 & 0 & 0 & 0 & 1 & 0
\end{pmatrix} . \label{cc}
\end{eqnarray}
Time-reversal symmetry is ${\mathcal T}^{\dagger} {\mathcal H}^{\ast} {\mathcal T} = {\mathcal H}$. For the position space Hamiltonian~(\ref{hpos}), the operator ${\mathcal T}$ is nonsymmorphic (NS), involving a translation,
\begin{eqnarray}
{\mathcal T} =
\begin{pmatrix}
0 & 0 & 1 & 0 & 0 & 0 & 0 & 0 & 0 & 0 & 0 & 0 \\
0 & 0 & 0 & 1 & 0 & 0 & 0 & 0 & 0 & 0 & 0 & 0 \\
0 & 0 & 0 & 0 & 1 & 0 & 0 & 0 & 0 & 0 & 0 & 0 \\
0 & 0 & 0 & 0 & 0 & 1 & 0 & 0 & 0 & 0 & 0 & 0 \\
0 & 0 & 0 & 0 & 0 & 0 & 1 & 0 & 0 & 0 & 0 & 0 \\
0 & 0 & 0 & 0 & 0 & 0 & 0 & 1 & 0 & 0 & 0 & 0 \\
0 & 0 & 0 & 0 & 0 & 0 & 0 & 0 & 1 & 0 & 0 & 0 \\
0 & 0 & 0 & 0 & 0 & 0 & 0 & 0 & 0 & 1 & 0 & 0 \\
0 & 0 & 0 & 0 & 0 & 0 & 0 & 0 & 0 & 0 & 1 & 0 \\
0 & 0 & 0 & 0 & 0 & 0 & 0 & 0 & 0 & 0 & 0 & 1 \\
1 & 0 & 0 & 0 & 0 & 0 & 0 & 0 & 0 & 0 & 0 & 0 \\
0 & 1 & 0 & 0 & 0 & 0 & 0 & 0 & 0 & 0 & 0 & 0
\end{pmatrix} . \label{tc}
\end{eqnarray}
Chiral symmetry is ${\mathcal S}^{\dagger} {\mathcal H} {\mathcal S} = -{\mathcal H}$. For the position space Hamiltonian~(\ref{hpos}), the operator ${\mathcal S}$ is NS, involving a translation,
where ${\mathcal S} = {\mathcal T}^{\ast} {\mathcal C}$,
\begin{eqnarray}
{\mathcal S} =
\begin{pmatrix}
0 & 0 & 0 & 1 & 0 & 0 & 0 & 0 & 0 & 0 & 0 & 0 \\
0 & 0 & 1 & 0 & 0 & 0 & 0 & 0 & 0 & 0 & 0 & 0 \\
0 & 0 & 0 & 0 & 0 & 1 & 0 & 0 & 0 & 0 & 0 & 0 \\
0 & 0 & 0 & 0 & 1 & 0 & 0 & 0 & 0 & 0 & 0 & 0 \\
0 & 0 & 0 & 0 & 0 & 0 & 0 & 1 & 0 & 0 & 0 & 0 \\
0 & 0 & 0 & 0 & 0 & 0 & 1 & 0 & 0 & 0 & 0 & 0 \\
0 & 0 & 0 & 0 & 0 & 0 & 0 & 0 & 0 & 1 & 0 & 0 \\
0 & 0 & 0 & 0 & 0 & 0 & 0 & 0 & 1 & 0 & 0 & 0 \\
0 & 0 & 0 & 0 & 0 & 0 & 0 & 0 & 0 & 0 & 0 & 1 \\
0 & 0 & 0 & 0 & 0 & 0 & 0 & 0 & 0 & 0 & 1 & 0 \\
0 & 1 & 0 & 0 & 0 & 0 & 0 & 0 & 0 & 0 & 0 & 0 \\
1 & 0 & 0 & 0 & 0 & 0 & 0 & 0 & 0 & 0 & 0 & 0
\end{pmatrix} .
\end{eqnarray}
When applying NS time-reversal and NS chiral symmetry operators to the position space Hamiltonian~(\ref{hpos}) with open boundary conditions, there are errors at the boundary because the boundary breaks the NS symmetry.

\subsection{Relation to the Kitaev chain}

The Kitaev chain~\cite{kitaev01s} is given by
\begin{eqnarray}
\!\!\!\!\!\!{\cal H} =
\begin{pmatrix}
-\mu & 0 & t & \Delta e^{i\phi} & 0 & 0 & 0 & 0 & 0 & 0 & 0 & 0 \\
0 & \mu & -\Delta e^{-i\phi} & -t & 0 & 0 & 0 & 0 & 0 & 0 & 0 & 0 \\
t & -\Delta e^{i\phi} & -\mu & 0 & t & \Delta e^{i\phi} & 0 & 0 & 0 & 0 & 0 & 0 \\
\Delta e^{-i\phi} & -t & 0 & \mu & -\Delta e^{-i\phi} & -t & 0 & 0 & 0 & 0 & 0 & 0 \\
0 & 0 & t & -\Delta e^{i\phi} & -\mu & 0 & t & \Delta e^{i\phi} & 0 & 0 & 0 & 0 \\
0 & 0 & \Delta e^{-i\phi} & -t & 0 & \mu & -\Delta e^{-i\phi} & -t & 0 & 0 & 0 & 0 \\
0 & 0 & 0 & 0 & t & -\Delta e^{i\phi} & -\mu & 0 & t & \Delta e^{i\phi} & 0 & 0 \\
0 & 0 & 0 & 0 & \Delta e^{-i\phi} & -t & 0 & \mu & -\Delta e^{-i\phi} & -t & 0 & 0 \\
0 & 0 & 0 & 0 & 0 & 0 & t & -\Delta e^{i\phi} & -\mu & 0 & t & \Delta e^{i\phi} \\
0 & 0 & 0 & 0 & 0 & 0 & \Delta e^{-i\phi} & -t & 0 & \mu & -\Delta e^{-i\phi} & -t \\
0 & 0 & 0 & 0 & 0 & 0 & 0 & 0 & t & -\Delta e^{i\phi} & -\mu & 0 \\
0 & 0 & 0 & 0 & 0 & 0 & 0 & 0 & \Delta e^{-i\phi} & -t & 0 & \mu
\end{pmatrix} . \label{hkpos}
\end{eqnarray}
In this Hamiltonian, the phase $\phi$ can be gauged away.
The Kitaev model (with the phase $\phi$ gauged away) is in symmetry class BDI with a $\mathbb{Z}$ topological index, the winding number W. With only nearest-neighbor hopping, the winding number can take three values~\cite{spanslatt15s},
\begin{eqnarray}
W = \begin{cases}
			1 & \text{for}\,\,|\mu| < 2 |t| \,\,\text{and}\,\,t\Delta > 0\\
            0 & \text{for}\,\,|\mu| > 2 |t|\\
            -1 & \text{for}\,\,|\mu| < 2 |t| \,\,\text{and}\,\,t\Delta < 0
		 \end{cases} .
\end{eqnarray}

For the $\mathbb{Z}_4$ model in phases $\theta = 1$ and $\theta = 3$, then $\mu ^2 + 4\Delta^2 \sin^2 \phi < 4t^2$. Then we can set $\sin \phi = 0$ (i.e. $\phi = \pi$ or $\phi = 0$) without causing a phase transition, i.e. $\phi$ is real.
For example, if we set $\phi = 0$ in the $\mathbb{Z}_4$ Hamiltonian~(\ref{hpos}), it will be the same as the Kitaev model~(\ref{hkpos}) with the phase $\phi$ gauged away.

\subsection{Relation to the CDW model}

For phases $\theta = 2$ and $\theta = 4$, then $\mu^2 + 4\Delta^2 \sin^2 \phi > 4t^2$. Without causing a phase transition, we can then set $t=0$. Then, as long as $\Delta \neq 0$ and $\sin \phi \neq 0$, we could have a phase transition between $\theta = 2$ and $\theta = 4$ by flipping the sign of $\mu$ (via $\mu = 0$).
With $t = 0$, the $\mathbb{Z}_4$ Hamiltonian~(\ref{hpos}) is
\begin{eqnarray}
\!\!\!\!\!\!{\cal H} =
\begin{pmatrix}
-\mu & 0 & 0 & \Delta e^{i\phi} & 0 & 0 & 0 & 0 & 0 & 0 & 0 & 0 \\
0 & \mu & -\Delta e^{-i\phi} & 0 & 0 & 0 & 0 & 0 & 0 & 0 & 0 & 0 \\
0 & -\Delta e^{i\phi} & -\mu & 0 & 0 & \Delta e^{-i\phi} & 0 & 0 & 0 & 0 & 0 & 0 \\
\Delta e^{-i\phi} & 0 & 0 & \mu & -\Delta e^{i\phi} & 0 & 0 & 0 & 0 & 0 & 0 & 0 \\
0 & 0 & 0 & -\Delta e^{-i\phi} & -\mu & 0 & 0 & \Delta e^{i\phi} & 0 & 0 & 0 & 0 \\
0 & 0 & \Delta e^{i\phi} & 0 & 0 & \mu & -\Delta e^{-i\phi} & 0 & 0 & 0 & 0 & 0 \\
0 & 0 & 0 & 0 & 0 & -\Delta e^{i\phi} & -\mu & 0 & 0 & \Delta e^{-i\phi} & 0 & 0 \\
0 & 0 & 0 & 0 & \Delta e^{-i\phi} & 0 & 0 & \mu & -\Delta e^{i\phi} & 0 & 0 & 0 \\
0 & 0 & 0 & 0 & 0 & 0 & 0 & -\Delta e^{-i\phi} & -\mu & 0 & 0 & \Delta e^{i\phi} \\
0 & 0 & 0 & 0 & 0 & 0 & \Delta e^{i\phi} & 0 & 0 & \mu & -\Delta e^{-i\phi} & 0 \\
0 & 0 & 0 & 0 & 0 & 0 & 0 & 0 & 0 & -\Delta e^{i\phi} & -\mu & 0 \\
0 & 0 & 0 & 0 & 0 & 0 & 0 & 0 & \Delta e^{-i\phi} & 0 & 0 & \mu
\end{pmatrix} .
\end{eqnarray}
We change basis with ${\cal H}^{\prime} = \hat{U} {\cal H} \hat{U}^{\dagger}$ where
\begin{eqnarray}
\hat{U} = \begin{pmatrix}
1 & 0 & 0 & 0 & 0 & 0 & 0 & 0 & 0 & 0 & 0 & 0 \\
0 & 0 & 0 & 1 & 0 & 0 & 0 & 0 & 0 & 0 & 0 & 0 \\
0 & 0 & 0 & 0 & 1 & 0 & 0 & 0 & 0 & 0 & 0 & 0 \\
0 & 0 & 0 & 0 & 0 & 0 & 0 & 1 & 0 & 0 & 0 & 0 \\
0 & 0 & 0 & 0 & 0 & 0 & 0 & 0 & 1 & 0 & 0 & 0 \\
0 & 0 & 0 & 0 & 0 & 0 & 0 & 0 & 0 & 0 & 0 & 1 \\
0 & 1 & 0 & 0 & 0 & 0 & 0 & 0 & 0 & 0 & 0 & 0 \\
0 & 0 & 1 & 0 & 0 & 0 & 0 & 0 & 0 & 0 & 0 & 0 \\
0 & 0 & 0 & 0 & 0 & 1 & 0 & 0 & 0 & 0 & 0 & 0 \\
0 & 0 & 0 & 0 & 0 & 0 & 1 & 0 & 0 & 0 & 0 & 0 \\
0 & 0 & 0 & 0 & 0 & 0 & 0 & 0 & 0 & 1 & 0 & 0 \\
0 & 0 & 0 & 0 & 0 & 0 & 0 & 0 & 0 & 0 & 1 & 0
\end{pmatrix} ,
\end{eqnarray}
giving
\begin{eqnarray}
\!\!\!\!\!\! \!\!\!\!\!\!{\cal H}^{\prime} = \begin{pmatrix}
-\mu & \Delta e^{i\phi} & 0 & 0 & 0 & 0 & 0 & 0 & 0 & 0 & 0 & 0 \\
\Delta e^{-i\phi} & \mu & -\Delta e^{i\phi} & 0 & 0 & 0 & 0 & 0 & 0 & 0 & 0 & 0 \\
0 & -\Delta e^{-i\phi} & -\mu & \Delta e^{i\phi} & 0 & 0 & 0 & 0 & 0 & 0 & 0 & 0 \\
0 & 0 & \Delta e^{-i\phi} & \mu & -\Delta e^{i\phi} & 0 & 0 & 0 & 0 & 0 & 0 & 0 \\
0 & 0 & 0 & -\Delta e^{-i\phi} & -\mu & \Delta e^{i\phi} & 0 & 0 & 0 & 0 & 0 & 0 \\
0 & 0 & 0 & 0 & \Delta e^{-i\phi} & \mu & 0 & 0 & 0 & 0 & 0 & 0 \\
0 & 0 & 0 & 0 & 0 & 0 & \mu & -\Delta e^{-i\phi} & 0 & 0 & 0 & 0 \\
0 & 0 & 0 & 0 & 0 & 0 & -\Delta e^{i\phi} & -\mu & \Delta e^{-i\phi} & 0 & 0 & 0 \\
0 & 0 & 0 & 0 & 0 & 0 & 0 & \Delta e^{i\phi} & \mu & -\Delta e^{-i\phi} & 0 & 0 \\
0 & 0 & 0 & 0 & 0 & 0 & 0 & 0 & -\Delta e^{i\phi} & -\mu & \Delta e^{-i\phi} & 0 \\
0 & 0 & 0 & 0 & 0 & 0 & 0 & 0 & 0 & \Delta e^{i\phi} & \mu & -\Delta e^{-i\phi} \\
0 & 0 & 0 & 0 & 0 & 0 & 0 & 0 & 0 & 0 & -\Delta e^{i\phi} & -\mu
\end{pmatrix} ,
\end{eqnarray}
It is possible to gauge away the phases of $\Delta$ with ${\cal H}^{\prime\prime} = R {\cal H}^{\prime} R^{\dagger}$ where
\begin{eqnarray}
R = \begin{pmatrix}
1 & 0 & 0 & 0 & 0 & 0 & 0 & 0 & 0 & 0 & 0 & 0 \\
0 & e^{i\phi} & 0 & 0 & 0 & 0 & 0 & 0 & 0 & 0 & 0 & 0 \\
0 & 0 & -e^{2i\phi} & 0 & 0 & 0 & 0 & 0 & 0 & 0 & 0 & 0 \\
0 & 0 & 0 & -e^{3i\phi} & 0 & 0 & 0 & 0 & 0 & 0 & 0 & 0 \\
0 & 0 & 0 & 0 & e^{4i\phi} & 0 & 0 & 0 & 0 & 0 & 0 & 0 \\
0 & 0 & 0 & 0 & 0 & e^{5i\phi} & 0 & 0 & 0 & 0 & 0 & 0 \\
0 & 0 & 0 & 0 & 0 & 0 & 1 & 0 & 0 & 0 & 0 & 0 \\
0 & 0 & 0 & 0 & 0 & 0 & 0 & -e^{-i\phi} & 0 & 0 & 0 & 0 \\
0 & 0 & 0 & 0 & 0 & 0 & 0 & 0 & -e^{-2i\phi} & 0 & 0 & 0 \\
0 & 0 & 0 & 0 & 0 & 0 & 0 & 0 & 0 & e^{-3i\phi} & 0 & 0 \\
0 & 0 & 0 & 0 & 0 & 0 & 0 & 0 & 0 & 0 & e^{-4i\phi} & 0 \\
0 & 0 & 0 & 0 & 0 & 0 & 0 & 0 & 0 & 0 & 0 & -e^{-5i\phi}
\end{pmatrix} ,
\end{eqnarray}
so that
\begin{eqnarray}
{\cal H}^{\prime\prime} = \begin{pmatrix}
-\mu & \Delta & 0 & 0 & 0 & 0 & 0 & 0 & 0 & 0 & 0 & 0 \\
\Delta & \mu & \Delta & 0 & 0 & 0 & 0 & 0 & 0 & 0 & 0 & 0 \\
0 & \Delta & -\mu & \Delta & 0 & 0 & 0 & 0 & 0 & 0 & 0 & 0 \\
0 & 0 & \Delta & \mu & \Delta & 0 & 0 & 0 & 0 & 0 & 0 & 0 \\
0 & 0 & 0 & \Delta & -\mu & \Delta & 0 & 0 & 0 & 0 & 0 & 0 \\
0 & 0 & 0 & 0 & \Delta & \mu & 0 & 0 & 0 & 0 & 0 & 0 \\
0 & 0 & 0 & 0 & 0 & 0 & \mu & \Delta & 0 & 0 & 0 & 0 \\
0 & 0 & 0 & 0 & 0 & 0 & \Delta & -\mu & \Delta & 0 & 0 & 0 \\
0 & 0 & 0 & 0 & 0 & 0 & 0 & \Delta & \mu & \Delta & 0 & 0 \\
0 & 0 & 0 & 0 & 0 & 0 & 0 & 0 & \Delta & -\mu & \Delta & 0 \\
0 & 0 & 0 & 0 & 0 & 0 & 0 & 0 & 0 & \Delta & \mu & \Delta \\
0 & 0 & 0 & 0 & 0 & 0 & 0 & 0 & 0 & 0 &-\Delta & -\mu
\end{pmatrix} .
\end{eqnarray}
This is block diagonal, effectively describing two separate CDW models with nearest-neighbor hopping $\Delta$ and alternating onsite energies $\pm \mu$.
Each of them is in a $\mathbb{Z}_2$ phase~\cite{shiozaki15s,brzezicki20s,mccann23s} depending on the sign of $\mu$.
We define the topological index $\nu$ for the CDW chain with the first site of onsite energy $-\mu$ as 
\begin{eqnarray*}
\nu =
\begin{cases}
0 & \text{for}\,\,\mu < 0 \\
1 & \text{for}\,\,\mu > 0
\end{cases}
\end{eqnarray*}
Hence phase $\theta = 4$ of the $\mathbb{Z}_4$ model corresponds to $\nu = 1$, and $\theta = 2$ corresponds to $\nu = 0$.

\section{Bogoliubov de Gennes representation in $k$ space: Type II tight-binding model}

There are two different possible choices of Fourier transform~\cite{bena09s}, either one can use lattice vectors (Type I) or orbital positions (Type II). For describing NS symmetries, we find it more convenient to use the Type II representation, as in the main text, because the NS symmetry operators are generally dependent on $k$ in the Type I representation, but not in the Type II representation. The two representations are related by a unitary transformation. Here we express the Hamiltonian and the symmetry operators in both representations.

\subsection{Hamiltonian}

We Fourier transform creation and annihilation operators using
\begin{eqnarray}
c_{n,\alpha} &=& \frac{1}{\sqrt{N}} \sum_k c_{k\alpha} e^{ikna} , \qquad \qquad
c_{n,\alpha}^{\dagger} = \frac{1}{\sqrt{N}} \sum_k c_{k\alpha}^{\dagger} e^{-ikna} , \label{ca2} \\
c_{n,\beta} &=& \frac{1}{\sqrt{N}} \sum_k c_{k\beta} e^{ikna + ika/2} , \qquad \qquad
c_{n,\beta}^{\dagger} = \frac{1}{\sqrt{N}} \sum_k c_{k\beta}^{\dagger} e^{-ikna - ika/2} , \label{cb2}
\end{eqnarray}
where $N$ is the number of unit cells and $a$ is the lattice constant.
The Hamiltonian~(\ref{sh1}) may be written in the BdG representation as $H = \frac{1}{2} \sum_k \Psi_k^{\dagger} {\cal H} (k) \Psi_k + \frac{1}{2} \sum_k \mathrm{tr} (h(k))$ where $\Psi_k = \begin{pmatrix} c_{k \alpha} &  c_{k \beta} & c_{-k \alpha}^{\dagger} & c_{-k \beta}^{\dagger} \end{pmatrix}^T$ for wave vector $k$ and the $4 \times 4$ Bloch Hamiltonian ${\cal H} (k)$ is
\begin{eqnarray}
{\cal H} (k) = \begin{pmatrix}
-\mu & 2t \cos (ka/2) & 0 & 2i \Delta \sin(ka/2 + \phi) \\
2t \cos (ka/2) & -\mu & 2i \Delta \sin(ka/2 - \phi) & 0 \\
0 & - 2i \Delta \sin(ka/2 - \phi) & \mu & - 2t \cos (ka/2) \\
-2i \Delta \sin(ka/2 + \phi) & 0 & - 2t \cos (ka/2) & \mu
\end{pmatrix} . \label{hk2}
\end{eqnarray}

\subsection{Symmetry operators}

The Bloch Hamiltonian~(\ref{hk2}) incorporates particle-hole symmetry as
\begin{eqnarray}
\text{particle-hole:} \qquad 
C^{\dagger} {\cal H}^{\ast} (k) C = - {\cal H} (-k) ,
\end{eqnarray}
where $C = \tau_x \sigma_0$ is a $4 \times 4$ matrix,
\begin{eqnarray}
C = \begin{pmatrix}
0 & 0 & 1 & 0 \\
0 & 0 & 0 & 1 \\
1 & 0 & 0 & 0 \\
0 & 1 & 0 & 0
\end{pmatrix} .
\end{eqnarray}
The Hamiltonian~(\ref{hk2}) breaks time-reversal symmetry, but the NS time-reversal symmetry acts as
\begin{eqnarray}
\text{NS\,\,time:} \qquad 
T^{\dagger} {\cal H}^{\ast} (k) T = {\cal H} (-k) , \label{nstimek2}
\end{eqnarray}
where $T = \tau_0 \sigma_x$ is a $4 \times 4$ matrix,
\begin{eqnarray}
T = \begin{pmatrix}
0 & 1 & 0 & 0 \\
1 & 0 & 0 & 0 \\
0 & 0 & 0 & 1 \\
0 & 0 & 1 & 0
\end{pmatrix} . 
\end{eqnarray}
The combination of particle-hole symmetry and NS time-reversal symmetry yields NS chiral symmetry,
\begin{eqnarray}
\text{NS\,\,chiral:} \qquad 
S^{\dagger} {\cal H} (k) S = - {\cal H} (k) , \label{nschiralk2}
\end{eqnarray}
where $S = \tau_x \sigma_x$ is a $4 \times 4$ matrix,
\begin{eqnarray}
S = \begin{pmatrix}
0 & 0 & 0 & 1 \\
0 & 0 & 1 & 0 \\
0 & 1 & 0 & 0 \\
1 & 0 & 0 & 0
\end{pmatrix} . 
\end{eqnarray}

\section{Bogoliubov de Gennes representation in $k$ space: Type I tight-binding model}

\subsection{Hamiltonian}

Instead of using Eqs.~(\ref{ca2},\ref{cb2}), we Fourier transform creation and annihilation operators using
\begin{eqnarray}
c_{n,\alpha} &=& \frac{1}{\sqrt{N}} \sum_k c_{k\alpha} e^{ikna} , \qquad \qquad
c_{n,\alpha}^{\dagger} = \frac{1}{\sqrt{N}} \sum_k c_{k\alpha}^{\dagger} e^{-ikna} , \label{ca1} \\
c_{n,\beta} &=& \frac{1}{\sqrt{N}} \sum_k c_{k\beta} e^{ikna} , \qquad \qquad
c_{n,\beta}^{\dagger} = \frac{1}{\sqrt{N}} \sum_k c_{k\beta}^{\dagger} e^{-ikna} , \label{cb1}
\end{eqnarray}
where $N$ is the number of unit cells and $a$ is the lattice constant.
Then the $4 \times 4$ Bloch Hamiltonian ${\cal \tilde{H}} (k)$ is
\begin{eqnarray}
{\cal \tilde{H}} (k) = \begin{pmatrix}
-\mu & t (1 + e^{-ika}) & 0 & \Delta \big( e^{i\phi} - e^{-i\phi - ika} \big) \\
t (1 + e^{ika}) & -\mu & - \Delta \big( e^{i\phi} - e^{-i\phi + ika} \big) & 0 \\
0 & - \Delta \big( e^{-i\phi} - e^{i\phi - ika} \big) & \mu & - t (1 + e^{-ika}) \\
\Delta \big( e^{-i\phi} - e^{i\phi + ika} \big) & 0 & - t (1 + e^{ika}) & \mu
\end{pmatrix} . \label{hk1}
\end{eqnarray}
This is related to ${\cal H} (k)$~(\ref{hk2}) by a unitary transformation ${\cal \tilde{H}} (k) = \tilde{U}_k {\cal H} (k) \tilde{U}_k^{\dagger}$ where
\begin{eqnarray}
\tilde{U}_k = \begin{pmatrix}
1 & 0 & 0 & 0 \\
0 & e^{ika/2} & 0 & 0 \\
0 & 0 & 1 & 0 \\
0 & 0 & 0 & e^{ika/2}
\end{pmatrix} . 
\end{eqnarray}

\subsection{Symmetry operators}

The Bloch Hamiltonian~(\ref{hk1}) incorporates particle-hole symmetry as
\begin{eqnarray}
\text{particle-hole:} \qquad 
\tilde{C}^{\dagger} {\cal \tilde{H}}^{\ast} (k) \tilde{C} = - {\cal \tilde{H}} (-k) ,
\end{eqnarray}
where $\tilde{C} = \tau_x \sigma_0$ is a $4 \times 4$ matrix,
\begin{eqnarray}
\tilde{C} = \begin{pmatrix}
0 & 0 & 1 & 0 \\
0 & 0 & 0 & 1 \\
1 & 0 & 0 & 0 \\
0 & 1 & 0 & 0
\end{pmatrix} .
\end{eqnarray}
The Hamiltonian~(\ref{hk1}) breaks time-reversal symmetry, but the NS time-reversal symmetry acts as
\begin{eqnarray}
\text{NS\,\,time:} \qquad 
\tilde{T}^{\dagger}(k) {\cal \tilde{H}}^{\ast} (k) \tilde{T}(k) = {\cal \tilde{H}} (-k) , \label{nstimek1}
\end{eqnarray}
where $\tilde{T}(k)$ is a $4 \times 4$ matrix,
\begin{eqnarray}
\tilde{T}(k) = \begin{pmatrix}
0 & e^{ika/2} & 0 & 0 \\
e^{-ika/2} & 0 & 0 & 0 \\
0 & 0 & 0 & e^{ika/2} \\
0 & 0 & e^{-ika/2} & 0
\end{pmatrix} . \label{type1nst}
\end{eqnarray}
The combination of particle-hole symmetry and NS time-reversal symmetry yields NS chiral symmetry,
\begin{eqnarray}
\text{NS\,\,chiral:} \qquad 
\tilde{S}^{\dagger}(k) {\cal \tilde{H}} (k) \tilde{S}(k) = - {\cal \tilde{H}} (k) , \label{nschiralk1}
\end{eqnarray}
where $\tilde{S}(k)$ is a $4 \times 4$ matrix,
\begin{eqnarray}
\tilde{S}(k) = \begin{pmatrix}
0 & 0 & 0 & e^{-ika/2} \\
0 & 0 & e^{ika/2} & 0 \\
0 & e^{-ika/2} & 0 & 0 \\
e^{ika/2} & 0 & 0 & 0
\end{pmatrix} . 
\end{eqnarray}

\section{Determination of the Majorana number}

To determine the Majorana number $M$~\cite{kitaev01s,budich13s,beenakker15s}, we use the Hamiltonian ${\cal \tilde{H}} (k)$~(\ref{hk1}) in the type I representation because it is $2\pi$ periodic in $ka$. We change the order of the basis ${\cal \bar{H}} (k) = \bar{U} {\cal \tilde{H}} (k) \bar{U}^{\dagger}$, where
\begin{eqnarray}
\bar{U} = \begin{pmatrix}
1 & 0 & 0 & 0 \\
0 & 0 & 1 & 0 \\
0 & 1 & 0 & 0 \\
0 & 0 & 0 & 1
\end{pmatrix} ,
\end{eqnarray}
giving
\begin{eqnarray}
{\cal \bar{H}} (k) = \begin{pmatrix}
-\mu & 0 & t (1 + e^{-ika}) & \Delta \big( e^{i\phi} - e^{-i\phi - ika} \big) \\
 0 & \mu & - \Delta \big( e^{-i\phi} - e^{i\phi - ika} \big) & -t (1 + e^{-ika}) \\
t (1 + e^{ika}) & - \Delta \big( e^{i\phi} - e^{-i\phi + ika} \big) & -\mu & 0 \\
\Delta \big( e^{-i\phi} - e^{i\phi + ika} \big) & - t (1 + e^{ika}) & 0 & \mu
\end{pmatrix} . 
\end{eqnarray}
Now we transform to the Majorana basis with $A(k) = -i \Omega {\cal \bar{H}} (k) \Omega^{\dagger}$~\cite{beenakker15s} where
\begin{eqnarray}
\Omega = \frac{1}{\sqrt{2}} \begin{pmatrix}
1 & 1 & 0 & 0 \\
i & - i & 0 & 0 \\
0 & 0 & 1 & 1 \\
0 & 0 & i & - i
\end{pmatrix} ,
\end{eqnarray}
and $A(k) =$
{\small\begin{eqnarray*}
\!\!\!\!\!\! \!\!\!\!\!\! \begin{pmatrix}
0 & \mu & \Delta \sin \phi (1 + e^{-ika}) & -t (1 + e^{-ika}) + \Delta \cos \phi (1 - e^{-ika}) \\
-\mu & 0 & t (1 + e^{-ika}) + \Delta \cos \phi (1 - e^{-ika}) & -\Delta \sin \phi (1 + e^{-ika}) \\
-\Delta \sin \phi (1 + e^{ika}) & - t (1 + e^{ika}) - \Delta \cos \phi (1 - e^{ika}) & 0 & \mu \\
t (1 + e^{ika}) - \Delta \cos \phi (1 - e^{ika}) & \Delta \sin \phi (1 + e^{ika}) & -\mu & 0
\end{pmatrix} \!\! ,
\end{eqnarray*}}
This is not an antisymmetric matrix for arbitrary $k$, but $A(0)$ and $A(\pi/a)$ are antisymmetric:
\begin{eqnarray}
A(0) =\begin{pmatrix}
0 & \mu & 2 \Delta \sin \phi & -2 t \\
-\mu & 0 & 2t & -2\Delta \sin \phi \\
-2\Delta \sin \phi & - 2t & 0 & \mu \\
2t & 2\Delta \sin \phi & -\mu & 0
\end{pmatrix} ,
\end{eqnarray}
\begin{eqnarray}
A(\pi/a) = \begin{pmatrix}
0 & \mu & 0 & 2\Delta \cos \phi \\
-\mu & 0 & 2\Delta \cos \phi & 0 \\
0 & - 2\Delta \cos \phi & 0 & \mu \\
- 2\Delta \cos \phi & 0 & -\mu & 0
\end{pmatrix} ,
\end{eqnarray}
The Majorana number~\cite{budich13s} is
\begin{eqnarray}
M &=& \mathrm{sgn} \big\{ \mathrm{Pf} [ A (0) ] \mathrm{Pf} [ A (\pi/a) ] \big\} , \nonumber \\
&=& \mathrm{sgn} \big\{ (\mu^2 + 4 \Delta^2 \sin^2 \phi - 4 t^2 ) (\mu^2 + 4 \Delta^2 \cos^2 \phi ) \big\} , \nonumber \\
&=& \mathrm{sgn} \big\{ \mu^2 + 4 \Delta^2 \sin^2 \phi - 4 t^2 \big\}  \qquad \text{for}\quad \mu^2 + 4 \Delta^2 \cos^2 \phi \neq 0 .
\end{eqnarray}

\section{Hamiltonian for a Josephson junction in a ring}

We follow~\cite{pientka13s} to describe a Josephson junction as a weak link in a ring enclosing magnetic flux $\Phi_B$, writing the $\mathbb{Z}_4$ Hamiltonian for $N$ unit cells as
\begin{eqnarray}
H &=& - \mu \sum_{n=1}^N \sum_{\sigma = \alpha , \beta} c_{n,\sigma}^{\dagger} c_{n,\sigma}
+ t \sum_{n=1}^{N} \left( c_{n,\alpha}^{\dagger} c_{n,\beta} + \mathrm{H.c.} \right)
+ t \sum_{n=1}^{N-1} \left( c_{n+1,\alpha}^{\dagger} c_{n,\beta} + \mathrm{H.c.} \right) \nonumber \\
&& + \sum_{n=1}^{N} \Big( \Delta e^{i \phi} c_{n,\alpha}^{\dagger} c_{n,\beta}^{\dagger} + \mathrm{H.c.} \Big)
+ \sum_{n=1}^{N-1} \Big( \Delta  e^{-i \phi} c_{n,\beta}^{\dagger} c_{n+1,\alpha}^{\dagger} + \mathrm{H.c.} \Big) + t^{\prime} \left( c_{1,\alpha}^{\dagger} c_{N,\beta} e^{- i \varphi_B} + \mathrm{H.c.} \right) ,
\end{eqnarray}
where $t^{\prime}$ is the strength of hopping across the weak link and the phase $\varphi_B$ is given by
\begin{eqnarray}
\varphi_B = \pi \frac{\Phi_B}{\Phi_{\mathrm{s}}} = \frac{e}{\hbar}\Phi_B ,
\end{eqnarray}
where $\Phi_{\mathrm{s}} = h/(2e)$ is the superconducting flux quantum.
For $N=3$ unit cells, the weak link modifies Eq.~(\ref{hpos}) as
\begin{eqnarray*}
{\cal H} =
\begin{pmatrix}
-\mu & 0 & t & \Delta e^{i\phi} & 0 & 0 & 0 & 0 & 0 & 0 & t^{\prime}e^{-i\varphi_B} & 0 \\
0 & \mu & -\Delta e^{-i\phi} & -t & 0 & 0 & 0 & 0 & 0 & 0 & 0 & -t^{\prime}e^{i\varphi_B} \\
t & -\Delta e^{i\phi} & -\mu & 0 & t & \Delta e^{-i\phi} & 0 & 0 & 0 & 0 & 0 & 0 \\
\Delta e^{-i\phi} & -t & 0 & \mu & -\Delta e^{i\phi} & -t & 0 & 0 & 0 & 0 & 0 & 0 \\
0 & 0 & t & -\Delta e^{-i\phi} & -\mu & 0 & t & \Delta e^{i\phi} & 0 & 0 & 0 & 0 \\
0 & 0 & \Delta e^{i\phi} & -t & 0 & \mu & -\Delta e^{-i\phi} & -t & 0 & 0 & 0 & 0 \\
0 & 0 & 0 & 0 & t & -\Delta e^{i\phi} & -\mu & 0 & t & \Delta e^{-i\phi} & 0 & 0 \\
0 & 0 & 0 & 0 & \Delta e^{-i\phi} & -t & 0 & \mu & -\Delta e^{i\phi} & -t & 0 & 0 \\
0 & 0 & 0 & 0 & 0 & 0 & t & -\Delta e^{-i\phi} & -\mu & 0 & t & \Delta e^{i\phi} \\
0 & 0 & 0 & 0 & 0 & 0 & \Delta e^{i\phi} & -t & 0 & \mu & -\Delta e^{-i\phi} & -t \\
t^{\prime}e^{i\varphi_B} & 0 & 0 & 0 & 0 & 0 & 0 & 0 & t & -\Delta e^{i\phi} & -\mu & 0 \\
0 & -t^{\prime}e^{-i\varphi_B} & 0 & 0 & 0 & 0 & 0 & 0 & \Delta e^{-i\phi} & -t & 0 & \mu
\end{pmatrix} . 
\end{eqnarray*}

\section{An arbitrary Hamiltonian in position space satisfying particle-hole and NS time-reversal symmetry}

\subsection{Form of an arbitrary Hamiltonian}

Since the nonsymmorphic symmetry involves a translation, tight-binding parameters must be constant throughout the whole system and, hence, the Hamiltonian is not robust in the presence of disorder~\cite{allen22s}.
We consider the form of an arbitrary Hermitian Hamiltonian in position space which preserves particle-hole symmetry and nonsymmorphic time-reversal symmetry. Including all possible terms (e.g., long-range hopping parameters), a $J \times J$ Hamiltonian is parameterized by only $\sim J$ real numbers~\cite{allen22s}.
A generic Hermitian Hamiltonian in position space satisying particle-hole symmetry~(\ref{cc}) and NS time-reversal symmetry~(\ref{tc}), when the number of orbitals $J$ is a multiple of $8$, may be written as
\begin{eqnarray}
{\mathcal H} =
\begin{pmatrix}
r_1 & 0 & {\color{blue} c_1} & {\color{blue} c_2} & {\color{blue} c_3} & {\color{blue} c_4} & {\color{blue} c_5} & {\color{blue} c_6} & f_1 & 0 & {\color{red} c_5} & {\color{red} -c_6^{\ast}} & {\color{red} c_3^{\ast}} & {\color{red} -c_4} & {\color{red} c_1} & {\color{red} -c_2^{\ast}} \\
0 & -r_1 & {\color{blue} -c_2^{\ast}} & {\color{blue} -c_1^{\ast}} & {\color{blue} -c_4^{\ast}} & {\color{blue} -c_3^{\ast}} & {\color{blue} -c_6^{\ast}} & {\color{blue} -c_5^{\ast}} & 0 & -f_1 & {\color{red} c_6} & {\color{red} -c_5^{\ast}} & {\color{red} c_4^{\ast}} & {\color{red} -c_3} & {\color{red} c_2} & {\color{red} -c_1^{\ast}} \\
{\color{red} c_1^{\ast}} & {\color{red} -c_2} & r_1 & 0 & {\color{blue} c_1^{\ast}} & {\color{blue} c_2^{\ast}} & {\color{blue} c_3^{\ast}} & {\color{blue} c_4^{\ast}} & {\color{blue} c_5^{\ast}} & {\color{blue} c_6^{\ast}} & f_1 & 0 & {\color{red} c_5^{\ast}} & {\color{red} -c_6} & {\color{red} c_3} & {\color{red} -c_4^{\ast}} \\
{\color{red} c_2^{\ast}} & {\color{red} -c_1} & 0 & -r_1 & {\color{blue} -c_2} & {\color{blue} -c_1} & {\color{blue} -c_4} & {\color{blue} -c_3} & {\color{blue} -c_6} & {\color{blue} -c_5} & 0 & -f_1 & {\color{red} c_6^{\ast}} & {\color{red} -c_5} & {\color{red} c_4} & {\color{red} -c_3^{\ast}} \\
{\color{red} c_3^{\ast}} & {\color{red} -c_4} & {\color{red} c_1} & {\color{red} -c_2^{\ast}} & r_1 & 0 & {\color{blue} c_1} & {\color{blue} c_2} & {\color{blue} c_3} & {\color{blue} c_4} & {\color{blue} c_5} & {\color{blue} c_6} & f_1 & 0 & {\color{red} c_5} & {\color{red} -c_6^{\ast}} \\
{\color{red} c_4^{\ast}} & {\color{red} -c_3} & {\color{red} c_2} & {\color{red} -c_1^{\ast}} & 0 & -r_1 & {\color{blue} -c_2^{\ast}} & {\color{blue} -c_1^{\ast}} & {\color{blue} -c_4^{\ast}} & {\color{blue} -c_3^{\ast}} & {\color{blue} -c_6^{\ast}} & {\color{blue} -c_5^{\ast}} & 0 & -f_1 & {\color{red} c_6} & {\color{red} -c_5^{\ast}} \\
{\color{red} c_5^{\ast}} & {\color{red} -c_6} & {\color{red} c_3} & {\color{red} -c_4^{\ast}} & {\color{red} c_1^{\ast}} & {\color{red} -c_2} & r_1 & 0 & {\color{blue} c_1^{\ast}} & {\color{blue} c_2^{\ast}} & {\color{blue} c_3^{\ast}} & {\color{blue} c_4^{\ast}} & {\color{blue} c_5^{\ast}} & {\color{blue} c_6^{\ast}} & f_1 & 0 \\
{\color{red} c_6^{\ast}} & {\color{red} -c_5} & {\color{red} c_4} & {\color{red} -c_3^{\ast}} & {\color{red} c_2^{\ast}} & {\color{red} -c_1} & 0 & -r_1 & {\color{blue} -c_2} & {\color{blue} -c_1} & {\color{blue} -c_4} & {\color{blue} -c_3} & {\color{blue} -c_6} & {\color{blue} -c_5} & 0 & -f_1 \\
f_1 & 0 & {\color{red} c_5} & {\color{red} -c_6^{\ast}} & {\color{red} c_3^{\ast}} & {\color{red} -c_4} & {\color{red} c_1} & {\color{red} -c_2^{\ast}} & r_1 & 0 & {\color{blue} c_1} & {\color{blue} c_2} & {\color{blue} c_3} & {\color{blue} c_4} & {\color{blue} c_5} & {\color{blue} c_6} \\
0 & -f_1 & {\color{red} c_6} & {\color{red} -c_5^{\ast}} & {\color{red} c_4^{\ast}} & {\color{red} -c_3} & {\color{red} c_2} & {\color{red} -c_1^{\ast}} & 0 & -r_1 & {\color{blue} -c_2^{\ast}} & {\color{blue} -c_1^{\ast}} & {\color{blue} -c_4^{\ast}} & {\color{blue} -c_3^{\ast}} & {\color{blue} -c_6^{\ast}} & {\color{blue} -c_5^{\ast}} \\
{\color{blue} c_5^{\ast}} & {\color{blue} c_6^{\ast}} & f_1 & 0 & {\color{red} c_5^{\ast}} & {\color{red} -c_6} & {\color{red} c_3} & {\color{red} -c_4^{\ast}} & {\color{red} c_1^{\ast}} & {\color{red} -c_2} & r_1 & 0 & {\color{blue} c_1^{\ast}} & {\color{blue} c_2^{\ast}} & {\color{blue} c_3^{\ast}} & {\color{blue} c_4^{\ast}} \\
{\color{blue} -c_6} & {\color{blue} -c_5} & 0 & -f_1 & {\color{red} c_6^{\ast}} & {\color{red} -c_5} & {\color{red} c_4} & {\color{red} -c_3^{\ast}} & {\color{red} c_2^{\ast}} & {\color{red} -c_1} & 0 & -r_1 & {\color{blue} -c_2} & {\color{blue} -c_1} & {\color{blue} -c_4} & {\color{blue} -c_3} \\
{\color{blue} c_3} & {\color{blue} c_4} & {\color{blue} c_5} & {\color{blue} c_6} & f_1 & 0 & {\color{red} c_5} & {\color{red} -c_6^{\ast}} & {\color{red} c_3^{\ast}} & {\color{red} -c_4} & {\color{red} c_1} & {\color{red} -c_2^{\ast}} & r_1 & 0 & {\color{blue} c_1} & {\color{blue} c_2} \\
{\color{blue} -c_4^{\ast}} & {\color{blue} -c_3^{\ast}} & {\color{blue} -c_6^{\ast}} & {\color{blue} -c_5^{\ast}} & 0 & -f_1 & {\color{red} c_6} & {\color{red} -c_5^{\ast}} & {\color{red} c_4^{\ast}} & {\color{red} -c_3} & {\color{red} c_2} & {\color{red} -c_1^{\ast}} & 0 & -r_1 & {\color{blue} -c_2^{\ast}} & {\color{blue} -c_1^{\ast}} \\
{\color{blue} c_1^{\ast}} & {\color{blue} c_2^{\ast}} & {\color{blue} c_3^{\ast}} & {\color{blue} c_4^{\ast}} & {\color{blue} c_5^{\ast}} & {\color{blue} c_6^{\ast}} & f_1 & 0 & {\color{red} c_5^{\ast}} & {\color{red} -c_6} & {\color{red} c_3} & {\color{red} -c_4^{\ast}} & {\color{red} c_1^{\ast}} & {\color{red} -c_2} & r_1 & 0 \\
{\color{blue} -c_2} & {\color{blue} -c_1} & {\color{blue} -c_4} & {\color{blue} -c_3} & {\color{blue} -c_6} & {\color{blue} -c_5} & 0 & -f_1 & {\color{red} c_6^{\ast}} & {\color{red} -c_5} & {\color{red} c_4} & {\color{red} -c_3^{\ast}} & {\color{red} c_2^{\ast}} & {\color{red} -c_1} & 0 & -r_1
\end{pmatrix} .
\end{eqnarray}
Here, the black matrix elements show $2 \times 2$ blocks of the form $r_1 \sigma_z$ or $f_1 \sigma_z$, where $r_1$ and $f_1$ are real. The blue matrix elements show $J/2 - 2$ complex numbers $c_i$, $i = 1,2,\ldots J/2-2$ which are repeated in a pattern with a period of $4$ rows, and periodicity means that this patterns continues into the bottom left part of the matrix. The red matrix elements are related to the blue ones by Hermicity of the matrix (hence they follow a pattern with a period of $4$ columns).

\subsection{Form of the Hamiltonian for $J=8$ orbitals}

We write out the form of the Hamiltonian explicitly for the simplest nontrivial case, with $J = 8$ orbitals.
Particle-hole symmetry is ${\mathcal C}^{\dagger} {\mathcal H}^{\ast} {\mathcal C} = -{\mathcal H}$ where
\begin{eqnarray}
{\mathcal C} =
\begin{pmatrix}
0 & 1 & 0 & 0 & 0 & 0 & 0 & 0 \\
1 & 0 & 0 & 0 & 0 & 0 & 0 & 0 \\
0 & 0 & 0 & 1 & 0 & 0 & 0 & 0 \\
0 & 0 & 1 & 0 & 0 & 0 & 0 & 0 \\
0 & 0 & 0 & 0 & 0 & 1 & 0 & 0 \\
0 & 0 & 0 & 0 & 1 & 0 & 0 & 0 \\
0 & 0 & 0 & 0 & 0 & 0 & 0 & 1 \\
0 & 0 & 0 & 0 & 0 & 0 & 1 & 0
\end{pmatrix} . 
\end{eqnarray}
A generic Hermitian Hamiltonian satisfying this may be written as
\begin{eqnarray}
{\mathcal H}_{C} =
\begin{pmatrix}
r_1 & 0 & c_1 & c_2 & c_3 & c_4 & c_5 & c_6 \\
0 & -r_1 & -c_2^{\ast} & -c_1^{\ast} & -c_4^{\ast} & -c_3^{\ast} & -c_6^{\ast} & -c_5^{\ast} \\
c_1^{\ast} & -c_2 & r_2 & 0 & c_7 & c_8 & c_9 & c_{10} \\
c_2^{\ast} & -c_1 & 0 & -r_2 & -c_8^{\ast} & -c_7^{\ast} & -c_{10}^{\ast} & -c_9^{\ast} \\
c_3^{\ast} & -c_4 & c_7^{\ast} & -c_8 & r_3 & 0 & c_{11} & c_{12} \\
c_4^{\ast} & -c_3 & c_8^{\ast} & -c_7 & 0 & -r_3 & -c_{12}^{\ast} & -c_{11}^{\ast} \\
c_5^{\ast} & -c_6 & c_9^{\ast} & -c_{10} & c_{11}^{\ast} & -c_{12} & r_4 & 0 \\
c_6^{\ast} & -c_5 & c_{10}^{\ast} & -c_9 & c_{12}^{\ast} & -c_{11} & 0 & -r_4
\end{pmatrix} , \label{c8}
\end{eqnarray}
where $r_i$, $i = 1,2,3,4$, are real and $c_j$, $j = 1,2,\ldots,12$ are complex.

Time-reversal symmetry is ${\mathcal T}^{\dagger} {\mathcal H}^{\ast} {\mathcal T} = {\mathcal H}$. For NS time-reversal symmetry, with a period of $2$ orbitals,
\begin{eqnarray}
{\mathcal T} =
\begin{pmatrix}
0 & 0 & 1 & 0 & 0 & 0 & 0 & 0 \\
0 & 0 & 0 & 1 & 0 & 0 & 0 & 0 \\
0 & 0 & 0 & 0 & 1 & 0 & 0 & 0 \\
0 & 0 & 0 & 0 & 0 & 1 & 0 & 0 \\
0 & 0 & 0 & 0 & 0 & 0 & 1 & 0 \\
0 & 0 & 0 & 0 & 0 & 0 & 0 & 1 \\
1 & 0 & 0 & 0 & 0 & 0 & 0 & 0 \\
0 & 1 & 0 & 0 & 0 & 0 & 0 & 0
\end{pmatrix} . \label{t8}
\end{eqnarray}
On applying this symmetry to Hamiltonian~(\ref{c8}), a generic Hamiltonian may be written as
\begin{eqnarray}
{\mathcal H}_{CT} =
\begin{pmatrix}
r_1 & 0 & c_1 & c_2 & f_1 & 0 & c_1 & -c_2^{\ast} \\
0 & -r_1 & -c_2^{\ast} & -c_1^{\ast} & 0 & -f_1 & c_2 & -c_1^{\ast} \\
c_1^{\ast} & -c_2 & r_1 & 0 & c_1^{\ast} & c_2^{\ast} & f_1 & 0 \\
c_2^{\ast} & -c_1 & 0 & -r_1 & -c_2 & -c_1 & 0 & -f_1 \\
f_1 & 0 & c_1 & -c_2^{\ast} & r_1 & 0 & c_1 & c_2 \\
0 & -f_1 & c_2 & -c_1^{\ast} & 0 & -r_1 & -c_2^{\ast} & -c_1^{\ast} \\
c_1^{\ast} & c_2^{\ast} & f_1 & 0 & c_1^{\ast} & -c_2 & r_1 & 0 \\
-c_2 & -c_1 & 0 & -f_1 & c_2^{\ast} & -c_1 & 0 & -r_1
\end{pmatrix} , \label{ct8}
\end{eqnarray}
where $r_1$, $f_1$ are real and $c_1$, $c_2$ are complex.

\subsection{Block diagonalization of the Hamiltonian with $J=8$ orbitals}

The NS time-reversal symmetry~(\ref{t8}) squares to give a unitary transformation ${\mathcal U}^{\dagger} {\mathcal H} {\mathcal U} = {\mathcal H}$ where
\begin{eqnarray}
{\mathcal U} =
\begin{pmatrix}
0 & 0 & 0 & 0 & 1 & 0 & 0 & 0 \\
0 & 0 & 0 & 0 & 0 & 1 & 0 & 0 \\
0 & 0 & 0 & 0 & 0 & 0 & 1 & 0 \\
0 & 0 & 0 & 0 & 0 & 0 & 0 & 1 \\
1 & 0 & 0 & 0 & 0 & 0 & 0 & 0 \\
0 & 1 & 0 & 0 & 0 & 0 & 0 & 0 \\
0 & 0 & 1 & 0 & 0 & 0 & 0 & 0 \\
0 & 0 & 0 & 1 & 0 & 0 & 0 & 0 \\
\end{pmatrix} .
\end{eqnarray}
We use its eigenstates to form an operator as
\begin{eqnarray}
{\mathcal R} = \frac{1}{\sqrt{2}}
\begin{pmatrix}
0 & 0 & 0 & 1 & 0 & 0 & 0 & -1 \\
0 & 1 & 0 & 0 & 0 & -1 & 0 & 0 \\
0 & 0 & 1 & 0 & 0 & 0 & -1 & 0 \\
1 & 0 & 0 & 0 & -1 & 0 & 0 & 0 \\
0 & 0 & 0 & 1 & 0 & 0 & 0 & 1 \\
0 & 1 & 0 & 0 & 0 & 1 & 0 & 0 \\
0 & 0 & 1 & 0 & 0 & 0 & 1 & 0 \\
1 & 0 & 0 & 0 & 1 & 0 & 0 & 0 \\
\end{pmatrix} .
\end{eqnarray}
and block diagonalize the Hamiltonian~(\ref{ct8}) as
${\mathcal H}_{B} = {\mathcal R}^{\dagger} {\mathcal H}_{CT} {\mathcal R}$ where
\begin{eqnarray}
{\mathcal H}_{B} =
\begin{pmatrix}
-u_1-f_1 & -2c_1& 0  & -2i \mathrm{Im}(c_2) & 0 & 0 & 0 & 0 \\
-2c_1^{\ast} & -u_1-f_1 & 2i \mathrm{Im}(c_2) & 0 & 0 & 0 & 0 & 0 \\
0 & -2i \mathrm{Im}(c_2) & u_1+f_1 & 2c_1^{\ast} & 0 & 0 & 0 & 0 \\
2i \mathrm{Im}(c_2) & 0 & 2c_1 & u_1+f_1 & 0 & 0 & 0 & 0 \\
0 & 0 & 0 & 0 & -u_1+f_1 & 0 & 0 & 2\mathrm{Re}(c_2) \\
0 & 0 & 0 & 0 & 0 & -u_1+f_1 & -2\mathrm{Re}(c_2) & 0 \\
0 & 0 & 0 & 0 & 0 & -2\mathrm{Re}(c_2) & u_1-f_1 & 0 \\
0 & 0 & 0 & 0 & 2\mathrm{Re}(c_2) & 0 & 0 & u_1-f_1 \\
\end{pmatrix} .
\end{eqnarray}
The upper left $4 \times 4$ block has time-reversal symmetry ${\mathcal T}_{1}^{\dagger} {\mathcal H}_{1}^{\ast} {\mathcal T}_{1} = {\mathcal H}_{1}$
where
\begin{eqnarray}
{\mathcal T}_{1} =
\begin{pmatrix}
0 & 1 & 0 & 0 \\
1 & 0 & 0 & 0 \\
0 & 0 & 0 & 1 \\
0 & 0 & 1 & 0
\end{pmatrix} , \label{t1b}
\end{eqnarray}
and the lower right $4 \times 4$ block has time-reversal symmetry ${\mathcal T}_{2}^{\dagger} {\mathcal H}_{2}^{\ast} {\mathcal T}_{2} = {\mathcal H}_{2}$
where
\begin{eqnarray}
{\mathcal T}_{2} =
\begin{pmatrix}
0 & i & 0 & 0 \\
-i & 0 & 0 & 0 \\
0 & 0 & 0 & i \\
0 & 0 & -i & 0
\end{pmatrix} . \label{t2b}
\end{eqnarray}
Here we have considered the simplest nontrivial case with $J=8$ orbitals, and the Hamiltonian block diagonalizes into two $4 \times 4$ blocks. One block corresponds to $k=0$ (with $\beta = 1$) and one block corresponds to $k = \pi /a$ (with $\beta = 4$). That each block (corresponding to a different $k$ value) has a different time-reversal symmetry may be easily seen by examining the form of the time-reversal symmetry operator in the Type I representation~(\ref{type1nst}),
\begin{eqnarray}
\tilde{T}(k) = \begin{pmatrix}
0 & e^{ika/2} & 0 & 0 \\
e^{-ika/2} & 0 & 0 & 0 \\
0 & 0 & 0 & e^{ika/2} \\
0 & 0 & e^{-ika/2} & 0
\end{pmatrix} .
\end{eqnarray}


\begin{thebibliography}{99}

\bibitem{altland97}
A. Altland and M. R. Zirnbauer,
Nonstandard symmetry classes in mesoscopic normal-superconducting hybrid structures,
Phys.\ Rev.\ B {\bf 55}, 1142 (1997).

\bibitem{schnyder08}
A. P. Schnyder, S. Ryu, A. Furusaki, and A. W. W. Ludwig,
Classification of topological insulators and superconductors in three spatial dimensions,
Phys.\ Rev.\ B {\bf 78}, 195125 (2008).

\bibitem{kitaev09}
A. Kitaev,
Periodic table for topological insulators and superconductors,
AIP Conf.\ Proc.\ {\bf 1134}, 22 (2009).

\bibitem{ryu10}
S. Ryu, A. P. Schnyder, A. Furusaki, and A. W. W. Ludwig,
Topological insulators and superconductors: tenfold way and dimensional hierarchy,
New J.\ Phys.\ {\bf 12}, 065010 (2010).

\bibitem{chiu16}
C.-K. Chiu, J. C. Y. Teo, A. P. Schnyder, and S. Ryu,
Classification of topological quantum matter with symmetries,
Rev.\ Mod.\ Phys.\ {\bf 88}, 035005 (2016).

\bibitem{liu14}
C.-X. Liu, R.-X. Zhang, and B. K. VanLeeuwen,
Topological nonsymmorphic crystalline insulators,
Phys.\ Rev.\ B {\bf 90}, 085304 (2014).

\bibitem{shiozaki14}
K. Shiozaki and M. Sato,
Topology of crystalline insulators and superconductors,
Phys.\ Rev.\ B {\bf 90}, 165114 (2014).

\bibitem{young15}
S. M. Young and C. L. Kane,
Dirac semimetals in two dimensions,
Phys.\ Rev.\ Lett.\ {\bf 115}, 126803 (2015).

\bibitem{wang16}
Z. Wang, A. Alexandradinata, R. J. Cava, and B. A. Bernevig,
Hourglass fermions,
Nature {\bf 532}, 189 (2016).

\bibitem{shiozaki16}
K. Shiozaki, M. Sato, and K. Gomi,
Topology of nonsymmorphic crystalline insulators and superconductors,
Phys.\ Rev.\ B {\bf 93}, 195413 (2016).

\bibitem{varjas17}
D. Varjas, F. de Juan, and Y.-M. Lu,
Space group constraints on weak indices in topological insulators,
Phys.\ Rev.\ B {\bf 96}, 035115 (2017).

\bibitem{kruthoff17}
J. Kruthoff, J. de Boer, J. van Wezel, C. L. Kane, and R.-J. Slager,
Topological classification of crystalline insulators through band structure combinatorics,
Phys.\ Rev.\ X {\bf 7}, 041069 (2017).

\bibitem{herrera22}
M. A. J. Herrera and D. Bercioux,
Tunable Dirac points in a two-dimensional nonsymmorphic wallpaper group lattice,
Commun.\ Phys.\ {\bf 6}, 42 (2023).

\bibitem{mong10}
R. S. K. Mong, A. M. Essin, and J. E. Moore,
Antiferromagnetic topological insulators,
Phys. Rev. B {\bf 81}, 245209 (2010).

\bibitem{fang15}
C. Fang and L. Fu,
New classes of three-dimensional topological crystalline insulators: Nonsymmorphic and magnetic,
Phys.\ Rev.\ B {\bf 91}, 161105(R) (2015).

\bibitem{shiozaki15}
K. Shiozaki, M. Sato, and K. Gomi,
$\mathbb{Z}_2$ topology in nonsymmorphic crystalline insulators: M\"obius twist in surface states,
Phys.\ Rev.\ B {\bf 91}, 155120 (2015).

\bibitem{zhao16}
Y. X. Zhao and A. P. Schnyder,
Nonsymmorphic symmetry-required band crossings in topological semimetals,
Phys.\ Rev.\ B {\bf 94}, 195109 (2016).

\bibitem{yanase17}
Y. Yanase and K. Shiozaki,
M\"obius topological superconductivity in UPt$_3$,
Phys. Rev. B {\bf 95}, 224514 (2017).

\bibitem{arkinstall17}
J. Arkinstall, M. H. Teimourpour, L. Feng, R. El-Ganainy, and H. Schomerus,
Topological tight-binding models from nontrivial square roots,
Phys.\ Rev.\ B {\bf 95}, 165109 (2017).

\bibitem{otrokov19}
M. M. Otrokov {\it et al.},
Prediction and observation of an antiferromagnetic topological insulator,
Nature {\bf 576}, 416 (2019).

\bibitem{gong19}
Y. Gong {\it et al.},
Experimental realization of an intrinsic magnetic topological insulator,
Chin.\ Phys.\ Lett.\ {\bf 36}, 076801 (2019).

\bibitem{zhang19}
D. Zhang, M. Shi, T. Zhu, D. Xing, H. Zhang, and J. Wang,
Topological axion states in the magnetic insulator MnBi$_2$Te$_4$ with the quantized magnetoelectric effect,
Phys.\ Rev.\ Lett.\ {\bf 122}, 206401 (2019).

\bibitem{niu20}
C. Niu, H. Wang, N. Mao, B. Huang, Y. Mokrousov, and Y. Dai,
Antiferromagnetic topological insulator with nonsymmorphic protection in two dimensions,
Phys.\ Rev.\ Lett.\ {\bf 124}, 066401 (2020).

\bibitem{marques19}
A. M. Marques and R. G. Dias,
One-dimensional topological insulators with noncentered inversion symmetry axis,
Phys.\ Rev.\ B {\bf 100}, 041104(R) (2019).

\bibitem{brzezicki20}
W. Brzezicki and T. Hyart,
Topological domain wall states in a nonsymmorphic chiral chain,
Phys.\ Rev.\ B {\bf 101}, 235113 (2020).

\bibitem{allen22}
R. E. J. Allen, H. V. Gibbons, A. M. Sherlock, H. R. M. Stanfield, and E. McCann,
Nonsymmorphic chiral symmetry and solitons in the Rice-Mele model,
Phys.\ Rev.\ B {\bf 106}, 165409 (2022).

\bibitem{yang22}
Y. Yang, H. C. Po, V. Liu, J. D. Joannopoulos, L. Fu, and M. Solja\v{c}i\'{c},
Non-Abelian nonsymmorphic chiral symmetries,
Phys.\ Rev.\ B {\bf 106}, L161108 (2022).

\bibitem{kivelson83}
S. Kivelson,
Solitons with adjustable charge in a commensurate Peierls insulator,
Phys.\ Rev.\ B {\bf 28}, 2653 (1983).

\bibitem{cayssol21}
J. Cayssol and J.-N. Fuchs,
Topological and geometrical aspects of band theory,
J.\ Phys.\ Mater.\ {\bf 4}, 034007 (2021).

\bibitem{fuchs21}
J.-N. Fuchs and F. Pi\'{e}chon,
Orbital embedding and topology of one-dimensional two-band insulators,
Phys.\ Rev.\ B {\bf 104}, 235428 (2021).

\bibitem{mccann23}
E. McCann,
Catalog of noninteracting tight-binding models with two energy bands in one dimension,
Phys.\ Rev.\ B {\bf 107}, 245401 (2023).

\bibitem{ezawa16}
M. Ezawa,
Hourglass fermion surface states in stacked topological insulators with nonsymmorphic symmetry,
Phys.\ Rev.\ B {\bf 94}, 155148 (2016).

\bibitem{chang17}
P.-Y. Chang, O. Erten, and P. Coleman,
M\"obius Kondo insulators,
Nat. Phys. {\bf 13}, 794 (2017).

\bibitem{daido19}
A. Daido, T. Yoshida, and Y. Yanase,
$\mathbb{Z}_4$ Topological Superconductivity in UCoGe,
Phys.\ Rev.\ Lett. {\bf 122}, 227001 (2019).

\bibitem{day23}
I. Araya Day, A. Varentcova, D. Varjas, and A. R. Akhmerov,
Pfaffian invariant identifies magnetic obstructed atomic insulators,
SciPost Phys. {\bf 15}, 114 (2023).

\bibitem{kitaev01}
A. Y. Kitaev,
Unpaired Majorana fermions in quantum wires,
Phys.-Usp.\ {\bf 44}, 131 (2001).

\bibitem{supplementary}
See Supplemental Material for details of the derivation of the BdG Hamiltonian in $k$ space, Eq.~(\ref{bdg1}), and position space, and the corresponding representation of the symmetry operators.

\bibitem{spanslatt15}
C. Sp\r{a}nsl\"att, E. Ardonne, J. C. Budich, and T. H. Hansson,
Topological aspects of $\pi$ phase winding junctions in superconducting wires,
J.\ Phys.: Condens.\ Matter {\bf 27}, 405701 (2015).

\bibitem{tanaka12}
Y. Tanaka, M. Sato, and N. Nagaosa,
Symmetry and Topology in Superconductors - Odd-Frequency Pairing and Edge States -,
J.\ Phys.\ Soc.\ Jpn.\ {\bf 81}, 011013 (2012).

\bibitem{alicea12}
J. Alicea,
New directions in the pursuit of Majorana fermions in solid state systems,
Rep.\ Prog.\ Phys.\ {\bf 75}, 076501 (2012).

\bibitem{leijnse12}
M. Leijnse and K. Flensberg,
Introduction to topological superconductivity and Majorana fermions,
Semicond.\ Sci.\ Technol.\ {\bf 27}, 124003 (2012).

\bibitem{beenakker15}
C. W. J. Beenakker, 
Random-matrix theory of Majorana fermions and topological superconductors,
Rev. Mod. Phys. {\bf 87}, 1037 (2015).

\bibitem{guo16}
H.-M. Guo,
A brief review on one-dimensional topological insulators and superconductors,
Sci.\ China Phys.\ Mech.\ Astron.\ {\bf 59}, 637401 (2016).

\bibitem{sato17}
M. Sato and Y. Ando,
Topological superconductors: a review,
Rep.\ Prog.\ Phys.\ {\bf 80}, 076501 (2017).

\bibitem{bena09}
C. Bena and G. Montambaux,
Remarks on the tight-binding model of graphene,
New J.\ Phys.\ {\bf 11}, 095003 (2009).

\bibitem{haim19}
A. Haim and Y. Oreg,
Time-reversal-invariant topological superconductivity in one and two dimensions,
Phys.\ Rep.\ {\bf 825}, 1 (2019).

\bibitem{bdgnote}
All plots are created for the BdG Hamiltonians ${\cal H}$ and a factor of $1/2$ should be included for the Hamiltonian $H = (1/2) \sum_k \Psi_k^{\dagger} {\cal H} (k) \Psi_k + \mathrm{const.}$

\bibitem{thetanote}
We slightly modify Eq.~(J35) in~\cite{shiozaki16} by replacing a natural logarithm with the arg function in order to describe the cases when the function $Z(k)$ is not unitary.

\bibitem{budich13}
J. C. Budich and E. Ardonne,
Equivalent topological invariants for one-dimensional Majorana wires in symmetry class D,
Phys.\ Rev.\ B {\bf 88}, 075419 (2013).

\bibitem{kwon04}
H.-J. Kwon, K. Sengupta, and V. M. Yakovenko,
Fractional ac Josephson effect in p- and d-wave superconductors,
Eur.\ Phys.\ J.\ B {\bf 37}, 349 (2004).

\bibitem{pientka13}
F. Pientka, A. Romito, M. Duckheim, Y. Oreg, and F. von Oppen,
Signatures of topological phase transitions in mesoscopic superconducting rings,
New J.\ Phys.\ {\bf 15}, 025001 (2013).

\bibitem{mehta90}
M. L. Mehta,
{\em Random Matrix Theory}
(Springer, New York, 1990)

\bibitem{guhr98}
T. Guhr, A. M\"uller-Groeling, and H. A. Weidenm\"uller,
Random-matrix theories in quantum physics: common concepts,
Phys.\ Rep.\ {\bf 299}, 189 (1998).

\bibitem{oganesyan07}
V. Oganesyan and D. A. Huse,
Localization of interacting fermions at high temperature,
Phys.\ Rev.\ B {\bf 75}, 155111 (2007).

\bibitem{atas13}
Y. Y. Atas, E. Bogomolny, O. Giraud, and G. Roux,
Distribution of the Ratio of Consecutive Level Spacings in Random Matrix Ensembles,
Phys.\ Rev.\ Lett. {\bf 110}, 084101 (2013).

\bibitem{lutchyn10}
R. M. Lutchyn, J. D. Sau, and S. Das Sarma,
Majorana Fermions and a Topological Phase Transition in Semiconductor-Superconductor Heterostructures,
Phys.\ Rev.\ Lett.\ {\bf 105}, 077001 (2010).

\bibitem{oreg10}
Y. Oreg, G. Refael, and F. von Oppen,
Helical Liquids and Majorana Bound States in Quantum Wires,
Phys.\ Rev.\ Lett.\ {\bf 105}, 177002 (2010).

\bibitem{prada20}
E. Prada, P. San-Jose, M. W. A. de Moor, A. Geresdi, E. J. H. Lee, J. Klinovaja, D. Loss, J. Nyg\r{a}rd, R. Aguado, and L. P. Kouwenhoven,
From Andreev to Majorana bound states in hybrid superconductor-semiconductor nanowires,
Nat.\ Rev.\ Phys.\ {\bf 2}, 575 (2020).

\bibitem{choy11}
T.-P. Choy, J. M. Edge, A. R. Akhmerov, and C. W. J. Beenakker,
Majorana fermions emerging from magnetic nanoparticles on a superconductor without spin-orbit coupling,
Phys.\ Rev.\ B {\bf 84}, 195442 (2011).

\bibitem{nadj-perge13}
S. Nadj-Perge, I. K. Drozdov, B. A. Bernevig, and A, Yazdani,
Proposal for realizing Majorana fermions in chains of magnetic atoms on a superconductor,
Phys.\ Rev.\ B {\bf 88}, 020407(R) (2013).

\bibitem{martin12}
I. Martin and A. F. Morpurgo,
Majorana fermions in superconducting helical magnets,
Phys.\ Rev.\ B {\bf 85}, 144505 (2012).

\bibitem{kjaergaard12}
M. Kjaergaard, K. W\"olms, and K. Flensberg,
Majorana fermions in superconducting nanowires without spin-orbit coupling,
Phys.\ Rev.\ B {\bf 85}, 020503(R) (2012).

\bibitem{hell17}
M. Hell, M. Leijnse, and K. Flensberg,
Two-Dimensional Platform for Networks of Majorana Bound States,
Phys.\ Rev.\ Lett.\ {\bf 118}, 107701 (2017).

\bibitem{pientka17}
F. Pientka, A. Keselman, E. Berg, A. Yacoby, A. Stern, and B. I. Halperin,
Topological Superconductivity in a Planar Josephson Junction,
Phys.\ Rev.\ X {\bf 7}, 021032 (2017).

\bibitem{sau12}
J. D. Sau and S. Das Sarma,
Realizing a robust practical Majorana chain in a quantum-dot-superconductor linear array,
Nat.\ Commun.\ {\bf 3}, 964 (2012).

\bibitem{leijnse12b}
M. Leijnse and K. Flensberg,
Parity qubits and poor man’s Majorana bound states in double quantum dots,
Phys.\ Rev.\ B {\bf 86}, 134528 (2012).

\bibitem{fulga13}
I. C. Fulga, A. Haim, A. R Akhmerov, and Y. Oreg,
Adaptive tuning of Majorana fermions in a quantum dot chain,
New J.\ Phys.\ {\bf 15}, 045020 (2013).

\bibitem{tsintzis24}
A. Tsintzis, R. S. Souto, K. Flensberg, J. Danon, and M. Leijnse,
Majorana Qubits and Non-Abelian Physics in Quantum Dot–Based Minimal
Kitaev Chains,
PRX quantum {\bf 5}, 010323 (2024).

\bibitem{lesser21a}
O. Lesser, K. Flensberg, F. von Oppen, and Y. Oreg,
Three-phase Majorana zero modes at tiny magnetic fields,
Phys.\ Rev.\ B {\bf 103}, L121116 (2021).

\bibitem{lesser21b}
O. Lesser, A. Saydjari, M. Wesson, A. Yacoby, and Y. Oreg,
Phase-induced topological superconductivity in a planar heterostructure,
PNAS {\bf 118}, e2107377118 (2021).

\bibitem{datanote}
\href{https://doi.org/10.17635/lancaster/researchdata/680}{https://doi.org/10.17635/lancaster/researchdata/680}

\end{thebibliography}

\begin{thebibliography}{99}

\bibitem{kitaev01s}
A. Y. Kitaev,
Unpaired Majorana fermions in quantum wires,
Phys.-Usp.\ {\bf 44}, 131 (2001).

\bibitem{spanslatt15s}
C. Sp\r{a}nsl\"att, E. Ardonne, J. C. Budich, and T. H. Hansson,
Topological aspects of $\pi$ phase winding junctions in superconducting wires,
J.\ Phys.: Condens.\ Matter {\bf 27}, 405701 (2015).

\bibitem{shiozaki15s}
K. Shiozaki, M. Sato, and K. Gomi,
$\mathbb{Z}_2$ topology in nonsymmorphic crystalline insulators: M\"obius twist in surface states,
Phys.\ Rev.\ B {\bf 91}, 155120 (2015).

\bibitem{brzezicki20s}
W. Brzezicki and T. Hyart,
Topological domain wall states in a nonsymmorphic chiral chain,
Phys.\ Rev.\ B {\bf 101}, 235113 (2020).

\bibitem{mccann23s}
E. McCann,
Catalog of noninteracting tight-binding models with two energy bands in one dimension,
Phys.\ Rev.\ B {\bf 107}, 245401 (2023).

\bibitem{bena09s}
C. Bena and G. Montambaux,
Remarks on the tight-binding model of graphene,
New J.\ Phys.\ {\bf 11}, 095003 (2009).

\bibitem{budich13s}
J. C. Budich and E. Ardonne,
Equivalent topological invariants for one-dimensional Majorana wires in symmetry class D,
Phys.\ Rev.\ B {\bf 88}, 075419 (2013).

\bibitem{beenakker15s}
C. W. J. Beenakker, 
Random-matrix theory of Majorana fermions and topological superconductors,
Rev. Mod. Phys. {\bf 87}, 1037 (2015).

\bibitem{pientka13s}
F. Pientka, A. Romito, M. Duckheim, Y. Oreg, and F. von Oppen,
Signatures of topological phase transitions in mesoscopic superconducting rings,
New J.\ Phys.\ {\bf 15}, 025001 (2013).

\bibitem{allen22s}
R. E. J. Allen, H. V. Gibbons, A. M. Sherlock, H. R. M. Stanfield, and E. McCann,
Nonsymmorphic chiral symmetry and solitons in the Rice-Mele model,
Phys.\ Rev.\ B {\bf 106}, 165409 (2022).

\end{thebibliography}
\end{document}